\documentclass[floats,floatfix,showpacs,amsmath,amssymb,aps,twocolumn,superscriptaddress,nofootinbib,nolongbibliography,prd]{revtex4-1}

\usepackage{amssymb,amsmath,verbatim,mathtools,needspace,enumitem,etoolbox,graphicx,physics,microtype,afterpage,xspace,tabularx,lmodern,multirow}
\newcommand\numberthis{\addtocounter{equation}{1}\tag{\theequation}}
\usepackage{gensymb}
\usepackage{soul}
\usepackage{multirow}
\usepackage{subcaption}
\usepackage{ragged2e}
\usepackage{bbm}
\usepackage[dvipsnames, usenames]{xcolor}
\definecolor{linkcolor}{rgb}{0.0,0.3,0.5}
\usepackage[unicode, colorlinks=true, linkcolor=linkcolor, citecolor=linkcolor, filecolor=linkcolor, urlcolor=linkcolor, linktocpage, breaklinks]{hyperref}
\usepackage[all]{hypcap}
\usepackage{cancel}
\usepackage[T1]{fontenc}
\usepackage[utf8]{inputenc}
\usepackage[usenames,dvipsnames]{xcolor}
\usepackage{csquotes}
\hypersetup{colorlinks=true,citecolor=Periwinkle,linkcolor=Periwinkle,urlcolor=Periwinkle}

\definecolor{romared}{RGB}{142,0,28}

\newcommand{\be}{\begin{equation}}
\newcommand{\ee}{\end{equation}}

\def\be{\begin{equation}}
\def\ee{\end{equation}}
\newcommand{\beq}{\begin{eqnarray}}
\newcommand{\eeq}{\end{eqnarray}}
\usepackage{makecell}
\usepackage{soul}

\newcolumntype{Y}{>{\centering\arraybackslash}X}

\newcommand{\bstheta}{\boldsymbol{\theta}}
\newcommand{\bspsi}{\boldsymbol{\psi}}
\newcommand{\bsvarphi}{\boldsymbol{\varphi}}

\makeatletter
\newcommand*{\rom}[1]{\expandafter\@slowromancap\romannumeral #1@}
\makeatother

\makeatletter
\let\jnl@style=\rm
\def\ref@jnl#1{{\jnl@style#1}}

\def\aj{\ref@jnl{AJ}}                   % Astronomical Journal
\def\actaa{\ref@jnl{Acta Astron.}}      % Acta Astronomica
\def\araa{\ref@jnl{ARA\&A}}             % Annual Review of Astron and Astrophys
\def\apj{\ref@jnl{ApJ}}                 % Astrophysical Journal
\def\apjl{\ref@jnl{ApJ}}                % Astrophysical Journal, Letters
\def\apjs{\ref@jnl{ApJS}}               % Astrophysical Journal, Supplement
\def\ao{\ref@jnl{Appl.~Opt.}}           % Applied Optics
\def\apss{\ref@jnl{Ap\&SS}}             % Astrophysics and Space Science
\def\aap{\ref@jnl{A\&A}}                % Astronomy and Astrophysics
\def\aapr{\ref@jnl{A\&A~Rev.}}          % Astronomy and Astrophysics Reviews
\def\aaps{\ref@jnl{A\&AS}}              % Astronomy and Astrophysics, Supplement
\def\azh{\ref@jnl{AZh}}                 % Astronomicheskii Zhurnal
\def\baas{\ref@jnl{BAAS}}               % Bulletin of the AAS
\def\bac{\ref@jnl{Bull. astr. Inst. Czechosl.}}
\def\caa{\ref@jnl{Chinese Astron. Astrophys.}}
\def\cjaa{\ref@jnl{Chinese J. Astron. Astrophys.}}
\def\icarus{\ref@jnl{Icarus}}           % Icarus
\def\jcap{\ref@jnl{J. Cosmology Astropart. Phys.}}
\def\jrasc{\ref@jnl{JRASC}}             % Journal of the RAS of Canada
\def\memras{\ref@jnl{MmRAS}}            % Memoirs of the RAS
\def\mnras{\ref@jnl{MNRAS}}             % Monthly Notices of the RAS
\def\na{\ref@jnl{New A}}                % New Astronomy
\def\nar{\ref@jnl{New A Rev.}}          % New Astronomy Review
\def\pra{\ref@jnl{Phys.~Rev.~A}}        % Physical Review A: General Physics
\def\prb{\ref@jnl{Phys.~Rev.~B}}        % Physical Review B: Solid State
\def\prc{\ref@jnl{Phys.~Rev.~C}}        % Physical Review C
\def\prd{\ref@jnl{Phys.~Rev.~D}}        % Physical Review D
\def\pre{\ref@jnl{Phys.~Rev.~E}}        % Physical Review E
\def\prl{\ref@jnl{Phys.~Rev.~Lett.}}    % Physical Review Letters
\def\pasa{\ref@jnl{PASA}}               % Publications of the Astron. Soc. of Australia
\def\pasp{\ref@jnl{PASP}}               % Publications of the ASP
\def\pasj{\ref@jnl{PASJ}}               % Publications of the ASJ
\def\rmxaa{\ref@jnl{Rev. Mexicana Astron. Astrofis.}}%
\def\qjras{\ref@jnl{QJRAS}}             % Quarterly Journal of the RAS
\def\skytel{\ref@jnl{S\&T}}             % Sky and Telescope
\def\solphys{\ref@jnl{Sol.~Phys.}}      % Solar Physics
\def\sovast{\ref@jnl{Soviet~Ast.}}      % Soviet Astronomy
\def\ssr{\ref@jnl{Space~Sci.~Rev.}}     % Space Science Reviews
\def\zap{\ref@jnl{ZAp}}                 % Zeitschrift fuer Astrophysik
\def\nat{\ref@jnl{Nature}}              % Nature
\def\iaucirc{\ref@jnl{IAU~Circ.}}       % IAU Cirulars
\def\aplett{\ref@jnl{Astrophys.~Lett.}} % Astrophysics Letters
\def\apspr{\ref@jnl{Astrophys.~Space~Phys.~Res.}}
\def\bain{\ref@jnl{Bull.~Astron.~Inst.~Netherlands}} 
\def\fcp{\ref@jnl{Fund.~Cosmic~Phys.}}  % Fundamental Cosmic Physics
\def\gca{\ref@jnl{Geochim.~Cosmochim.~Acta}}   % Geochimica Cosmochimica Acta
\def\grl{\ref@jnl{Geophys.~Res.~Lett.}} % Geophysics Research Letters
\def\jcp{\ref@jnl{J.~Chem.~Phys.}}      % Journal of Chemical Physics
\def\jgr{\ref@jnl{J.~Geophys.~Res.}}    % Journal of Geophysics Research
\def\jqsrt{\ref@jnl{J.~Quant.~Spec.~Radiat.~Transf.}}
\def\memsai{\ref@jnl{Mem.~Soc.~Astron.~Italiana}}
\def\nphysa{\ref@jnl{Nucl.~Phys.~A}}   % Nuclear Physics A
\def\physrep{\ref@jnl{Phys.~Rep.}}   % Physics Reports
\def\physscr{\ref@jnl{Phys.~Scr}}   % Physica Scripta
\def\planss{\ref@jnl{Planet.~Space~Sci.}}   % Planetary Space Science
\def\procspie{\ref@jnl{Proc.~SPIE}}   % Proceedings of the SPIE

\makeatother

\begin{document}

\title{Bias-Corrected Importance Sampling for Inferring Beyond-Vacuum-GR Effects in Gravitational-Wave Sources}

\author{Shubham Kejriwal} 
\email[]{shubhamkejriwal@u.nus.edu}
\affiliation{Department of Physics, National University of Singapore, Singapore 117551}
\author{Francisco Duque} 
\email[]{francisco.duque@aei.mpg.de}
\affiliation{Max Planck Institute for Gravitational Physics (Albert Einstein Institute), D-14476 Potsdam, Germany}
\author{Alvin J. K. Chua}
\email[]{alvincjk@nus.edu.sg}
\affiliation{Department of Physics, National University of Singapore, Singapore 117551}
\affiliation{Department of Mathematics, National University of Singapore, Singapore 119076}
\author{Jonathan Gair} 
\email[]{jonathan.gair@aei.mpg.de}
\affiliation{Max Planck Institute for Gravitational Physics (Albert Einstein Institute), D-14476 Potsdam, Germany}

\begin{abstract}
    The upcoming gravitational wave (GW) observatory LISA will measure the parameters of sources like extreme-mass-ratio inspirals (EMRIs) to exquisite precision. These measurements will also be sensitive to perturbations to the vacuum, GR-consistent evolution of sources, which might be caused by astrophysical environments or deviations from general relativity (GR). Previous studies have shown such ``beyond-vacuum-GR'' perturbations to potentially induce severe biases ($\gtrsim 10\sigma$) on recovered parameters under the ``null'' vacuum-GR hypothesis. While Bayesian inference can be performed under the null hypothesis using Markov Chain Monte Carlo (MCMC) samplers, it is computationally infeasible to repeat for more than a modest subset of all possible beyond-vacuum-GR hypotheses. We introduce bias-corrected importance sampling, a generic inference technique for nested models that is informed by the null hypothesis posteriors and the linear signal approximation to correct any induced inference biases. For a typical EMRI source that is significantly influenced by its environment but has been inferred only under the null hypothesis, the proposed method efficiently recovers the injected (unbiased) source parameters and the true posterior at a fraction of the expense of redoing MCMC inference under the full hypothesis. In future GW data analysis using the output of the proposed LISA global-fit pipeline, such methods may be necessary for the feasible and systematic inference of beyond-vacuum-GR effects. 
\end{abstract}

\maketitle 

%%%%%%%%%%%%%%%%%%%%%%%%%%%%%%%%%%%%%%%%%%%%%
\section{Introduction}\label{sec:intro}
%%%%%%%%%%%%%%%%%%%%%%%%%%%%%%%%%%%%%%%%%%%%%
The Laser Interferometer Space Antenna (LISA), recently adopted by the European Space Agency and scheduled to launch in the mid-2030s, is a space-based gravitational wave (GW) detector that will uncover new classes of GW sources emitting in the milli-Hz frequency band~\cite{LISA:2017pwj,LISA:2022yao,Colpi:2024xhw}. LISA is expected to tightly constrain the parameters characterising %model descriptions of 
sources like extreme-mass-ratio inspirals (EMRIs), in which a stellar-mass compact object (CO) of mass $\mu \sim 10^{1-2} M_\odot$ completes $\sim 10^5$ orbits around a supermassive black hole (MBH) of mass $M \sim 10^{4-7} M_\odot$ over a timescale of years~\cite{Barack:2003fp,Babak:2017tow,Berry:2019wgg}. Signals emitted by these strong-gravity sources evolving in matter-rich galactic nuclei would carry information about any potential deviations from GR as well as any astrophysical environments perturbing the binary's evolution, providing an exciting and unique opportunity to probe such ``beyond-vacuum-GR'' effects~\cite{Yunes:2009ke,Barausse:2016eii,LISA:2022kgy,Speri:2024qak,Barausse:2014tra,Speri:2022upm,Kocsis:2011dr,Garg:2024qxq,Duque:2024mfw,Favata:2013rwa,Romero-Shaw:2020thy,Romero-Shaw:2021ual,Piovano:2020zin,Mathews:2021rod,Skoupy:2024uan, Lyu:2024gnk}. 

The long timescale EMRI dynamics are modelled using black hole perturbation theory and self-force theory, in which the evolution of the system's conserved quantities is driven by its GW radiation and can be rewritten as a set of coupled ordinary differential equations known as the flux-balance laws~\cite{Hinderer:2008dm,Barack:2018yvs,Pound:2021qin,Fujita:2020zxe,Isoyama:2021jjd,Hughes:2021exa}. The contribution from a beyond-vacuum-GR effect is typically written as a perturbative addition to these laws as
\begin{align}
    \mathcal{F} = \mathcal{F}_{\rm GR}(\bspsi)(1 + \delta\mathcal{F}(\bspsi,\bsvarphi))\label{eq:perturbativebvgr}
\end{align}
where $\mathcal{F}_{\rm GR} := (\dot{E},\dot{L},\dot{\Phi}_\theta,\dot{\Phi}_r)$ in general, which are respectively the leading-order model-dependent fluxes of energy at infinity, the axial component of the angular momentum of the system, and the rate of change of the polar and azimuthal phase offsets, for the vacuum-GR parameters $\bspsi$. \footnote{Modifications from beyond-vacuum-GR effects to the phase fluxes vanish at the leading order, but are nevertheless described here for completeness.} $\delta{\mathcal F}$ is the effective beyond-vacuum-GR contribution, incorporating additional parameters $\bsvarphi$. 

If a true signal that is perturbed by a beyond-vacuum-GR effect is analyzed assuming the ``null'' vacuum-GR hypothesis (i.e., where $\bsvarphi = \bsvarphi_0$ is held fixed such that $\delta\mathcal{F} = 0$), the recovered best-fit estimate of $\bspsi$ is generally biased, possibly significantly, as shown in~\cite{Kejriwal:2023djc}. The schematic in the left panel of Fig. 1 geometrically visualizes the induced biases, where the injected signal (blue cross) generated in the full beyond-vacuum-GR hypothesis (red hypercube) is inferred in a null subset hypothesis (shaded submanifold). Correspondingly, the best-fit parameters are obtained at the orthogonal projection of the injection onto the restricted submanifold, introducing generic inference biases. See section~\ref{sec:basicimplementation} for more details. While Bayesian inference can be performed under the null hypothesis using techniques like Markov-Chain Monte Carlo (MCMC)~\cite{robert_monte_2004} and (dynamic) nested sampling~\cite{Skilling:2006gxv,2019S&C....29..891H}, it is computationally infeasible to repeat this procedure for more than a modest subset of all possible beyond-vacuum-GR hypotheses, which however is necessary for the robust interpretation of the signal. Instead, the vacuum-GR results and the perturbative beyond-vacuum-GR effects formulated in a nested-modeling framework can inform the inference of a competing hypothesis, through the bias-corrected importance sampling method formulated in this paper. In essence, our technique attempts to correct biases induced by a potential beyond-vacuum-GR effect present in the signal using the linear signal approximation for nested models~\cite{Cutler:2007mi,Kejriwal:2023djc}. Then, with random samples drawn from the corrected \textit{proposal} probability density function (pdf), we perform importance sampling to uncover the underlying posterior (i.e., the \textit{target} pdf). Although we illustrate the method by applying it to the problem of the inference of EMRI signals observed by LISA, the formalism we develop is generic and can be applied to any inference problem with nested models. Such problems arise quite generically in GW data analysis (see, e.g. \cite{Yunes:2009ke,Favata:2013rwa,LIGOScientific:2016lio,Romero-Shaw:2020thy,NANOGrav:2023pdq}).

After reviewing the prerequisites and defining our notation in the next section, we present in Section~\ref{sec:methods} two implementations within the same framework---a basic formalism conveying the method's main principles, and a more conservative regularized formalism which facilitates its practical application. We follow up with example results showcasing the method's effectiveness for LISA EMRIs in Section~\ref{sec:examples}, before discussing our outlook and future directions in Section~\ref{sec:discussion}.

%%%%%%%%%%%%%%%%%%%%%%%%%%%%%%%%%%%%%%%%%%%%%
\section{Background}\label{sec:background}
%%%%%%%%%%%%%%%%%%%%%%%%%%%%%%%%%%%%%%%%%%%%%

%%%%%%%%%%%%%%%%%%%%%%%%%%%%%%%%%%%%%%%%%%%%%
\subsection{Importance sampling}\label{sec:IS-methods}
%%%%%%%%%%%%%%%%%%%%%%%%%%%%%%%%%%%%%%%%%%%%
For a random variable $\bstheta \sim p(\bstheta)$, where $p(\bstheta)$ is a $d$-dimensional pdf, the expectation value of an arbitrary test function $\tau$ of $\bstheta$ is 
\begin{align}
    \mathbb{E}_{p}[\tau] = \int {\rm d}\bstheta~ \tau(\bstheta) p(\bstheta),
\end{align}
which requires the knowledge of $p(\bstheta)$, i.e., the target pdf. In scenarios where $p(\bstheta)$ is hard to sample from directly, especially true for more practical higher dimensional problems, the importance sampling method seeks to obtain $\mathbb{E}_{p}[\tau]$ by first randomly sampling from a simpler proposal pdf, $q(\bstheta)$, with a support that fully covers the target such that~\cite{gelmanbda04}
\begin{align*}
    \mathbb{E}_{p}[\tau] &= \int {\rm d}\bstheta~ \tau(\bstheta) p(\bstheta)\\
    &= \int {\rm d}\bstheta~ \tau(\bstheta)\frac{p(\bstheta)}{q(\bstheta)}q(\bstheta) = \mathbb{E}_{q}[\tau w]; \numberthis \label{eq:impexpect}\\
    w(\bstheta) &:= \frac{p(\bstheta)}{q(\bstheta)}.\numberthis
\end{align*}
When a finite set of samples $\{\bstheta^{(i)}\}_{i=1}^N \sim q(\bstheta)$ of size $N$ is available, Eq.~\eqref{eq:impexpect} can be approximated by the Monte Carlo integral~\cite{robert_monte_2004},
\begin{align}\label{eq:numericalexpect}
    \mathbb{E}_{p}[\tau] \approx \sum_{i=1}^N \tau(\bstheta_i)\hat{w}_i;~\hat{w}_i:= \frac{w(\bstheta_i)}{\sum_{i=1}^N w(\bstheta_i)} \, ,
\end{align}
and $\hat{w}_i$ is called the $i^{\rm th}$ sample's normalized \textit{importance weight}. This expression allows the target to be known only up to a normalization constant. Additionally, the distribution of random redraws from the set of samples $\{\bstheta^{(i)}\}_i$ chosen in proportion to their importance weights will align with the target pdf (in regions where the two overlap). This is known as sampling/importance resampling (SIR)~\cite{doi:https://doi.org/10.1002/0470090456.ch24} or simply importance resampling. The efficiency of importance sampling, $\eta$, can be quantified as
\begin{align}
    \eta &:= \frac{\hat{N}_{\rm ess}}{N}~\label{eq:impeff} \, , \\
    \hat{N}_{\rm ess} &= \frac{1}{\sum_{i=1}^N \hat{w}_i^2} \,,\\
    \mathbb{E}_{q}\left[ \hat{N}_{\rm ess} \right] &\approx\left(N{\rm var}[\hat{w}]+\frac{1}{N}\right)^{-1}\label{eq:ess}
\end{align}
where Eq.~\eqref{eq:ess} follows from the definition of variance on $\hat{w}$, ${\rm var}[\hat{w}]:= \mathbbm{E}_q[\hat{w}^2]-\mathbbm{E}_q[\hat{w}]^2$ and noting that $\mathbbm{E}_q[w]^2 = 1/N^2$.

Thus, the efficiency of importance sampling is maximized when ${\rm var}[\hat{w}]$ is minimized. This occurs as $q(\bstheta) \to p(\bstheta)$, in which limit the variance tends to zero. However, ensuring that $q(\bstheta) \approx p(\bstheta)$ requires knowledge of the target distribution which might not be available, and so this often does not hold, meaning that the efficiency drops dramatically in practice, particularly for higher-dimensional problems. Despite its simplistic formulation, perfect computational parallelizability, and virtually no arbitrary tuning parameters, importance sampling is thus considered inferior to sequential but dynamic sampling methods like Markov Chain Monte Carlo (MCMC)~\cite{robert_monte_2004} and (dynamic) nested sampling~\cite{Skilling:2006gxv,2019S&C....29..891H}. Yet, as we describe below, there exist scenarios in GW data analysis where the proposal and target distributions can be made similar, enabling its practical usage.

%%%%%%%%%%%%%%%%%%%%%%%%%%%%%%%%%%%%%%%%%%%%%
\subsection{GW data analysis and hypothesis testing}
%%%%%%%%%%%%%%%%%%%%%%%%%%%%%%%%%%%%%%%%%%%%%
In GW data analysis, we are usually interested in inferring the \textit{posterior} pdf $p(\bstheta|s,\mathcal{H})$, i.e., the probability distribution of the parameters $\bstheta$ described by a waveform template $h$ given the data $s$ and under the chosen hypothesis $\mathcal{H}$. Then, according to Bayes' theorem~\cite{jaynes03},
\begin{align}
    p(\bstheta|s,\mathcal{H}) &= \frac{\mathcal{L}(\bstheta|s,\mathcal{H})\pi(\bstheta|\mathcal{H})}{\mathcal{Z}(s|\mathcal{H})},\label{eq:bayestheorem}
\end{align}
where $\mathcal{L}(\bstheta|s,\mathcal{H}) := \exp\left[-0.5\left.\langle h(\bstheta)-s\right|h(\bstheta)-s\rangle\right]$ is the GW \textit{likelihood} function, $\pi(\bstheta|\mathcal{H})$ is the \textit{prior} pdf, and $\langle\cdot|\cdot\rangle$ is the canonical detector-noise-weighted inner product~\cite{Finn:1992wt,Cutler:1994ys}. $\mathcal{Z}$ is the \textit{evidence} for the data under the hypothesis, defined as the marginal likelihood
\begin{align}
    \mathcal{Z}(s|\mathcal{H}) &:= \int {\rm d}\bstheta~\mathcal{L}(\bstheta|s,\mathcal{H})\pi(\bstheta|\mathcal{H}).\label{eq:evidence}
\end{align}
While $s$ is generically the sum of the GW signal and detector noise, in the following text, we work in the zero-noise realization for simplicity.

The point $\bstheta_{\rm MAP}|\mathcal{H}$, where $p(\bstheta|s,\mathcal{H})$ takes its maximum value, is called the maximum a posteriori estimate (MAP) (which is also the maximum likelihood estimate in the flat-prior case). The conditioning of the MAP on the hypothesis is left explicit to stress that both the MAP and the posterior itself generally change under different hypotheses. In such scenarios, two competing hypotheses $\mathcal{H}_0$ and $\mathcal{H}_1$ can be compared using their posterior odds
\begin{align}
    \frac{p(\mathcal{H}_1|s)}{p(\mathcal{H}_0|s)} = \mathcal{B}_0^1 \times\frac{\pi(\mathcal{H}_1)}{\pi(\mathcal{H}_0)}
\end{align}
which reduce to the Bayes factor,
\begin{align}
    \mathcal{B}_0^1 &:= \frac{\mathcal{Z}(s|\mathcal{H}_1)}{\mathcal{Z}(s|\mathcal{H}_0)}. \label{eq:generalBF}
\end{align}
when we set equal priors on $\mathcal{H}_0$ and $\mathcal{H}_1$.

%%%%%%%%%%%%%%%%%%%%%%%%%%%%%%%%%%%%%%%%%%%%%
\subsection{Hypothesis testing in nested models}
%%%%%%%%%%%%%%%%%%%%%%%%%%%%%%%%%%%%%%%%%%%%%

The Bayes factor is extensively used in GW data analysis to test an alternate \textit{superset} hypothesis, $\mathcal{H}_1$, that includes a set of extra parameters $\bsvarphi$ of size $d_{\bsvarphi}$ that modify the null \textit{subset} hypothesis, $\mathcal{H}_0$, described by some common model parameters $\bspsi$ of size $d_{\bspsi}$~\cite{LIGOScientific:2016lio,LIGOScientific:2019fpa,LIGOScientific:2020tif,LIGOScientific:2021sio,NANOGrav:2023pdq,NANOGrav:2023gor,NANOGrav:2023icp,EuropeanPulsarTimingArray:2023egv,EPTA:2023fyk,EuropeanPulsarTimingArray:2023lqe,Favata:2013rwa,Payne:2019wmy,Romero-Shaw:2020thy,Romero-Shaw:2021ual}. The two hypotheses are equal when $\bsvarphi$ assumes a null value $\bsvarphi_0$ in $\mathcal{H}_1$. Our notation is summarized in Table~\ref{tab:notation}. 

Decomposing the parameter vector as $\bstheta := (\bspsi,\bsvarphi)$ of size $d:=d_{\bspsi}+d_{\bsvarphi}$, and denoting the likelihood and the prior in the superset hypothesis as $\mathcal{L}(\bspsi,\bsvarphi|s,\mathcal{H}_1)$ and $\pi(\bspsi,\bsvarphi|\mathcal{H}_1)$, the respective quantities in the subset hypothesis can be rewritten as the conditionals
\begin{align*}
    \mathcal{L}(\bspsi|s,\mathcal{H}_0) &= \mathcal{L}(\bspsi|\bsvarphi=\bsvarphi_0,s,\mathcal{H}_1),\label{eq:nestedlikelihood}\numberthis\\
    \pi(\bspsi|\mathcal{H}_0) &= \pi(\bspsi|\bsvarphi=\bsvarphi_0,\mathcal{H}_1)\\
    &= \pi(\bspsi|\mathcal{H}_1).\numberthis\label{eq:independentpriors}
\end{align*}
In Eq.~\eqref{eq:independentpriors} and the following, we assume that the priors on $\bspsi$ and $\bsvarphi$ are independent. Marginalising the posterior under $\mathcal{H}_1$ over $\bspsi$, and using Eqs.~\eqref{eq:bayestheorem},~\eqref{eq:nestedlikelihood}, and \eqref{eq:independentpriors}, we have $\pi(\bsvarphi=\bsvarphi_0|\mathcal{H}_1) \mathcal{Z}(s|\mathcal{H}_0) = \mathcal{Z}(s|\mathcal{H}_1) p(\bsvarphi=\bsvarphi_0|s,\mathcal{H}_1)$, where $p(\bsvarphi=\bsvarphi_0|s,\mathcal{H}_1)$ is the posterior pdf in the superset hypothesis marginalized over $\bspsi$ and evaluated at the null value of the extra parameter $\bsvarphi = \bsvarphi_0$. We then see that the Bayes factor reduces to the Savage-Dickey ratio for nested models~\cite{Dickey1971TheWL},
\begin{align}
    \mathcal{B}_{0}^1 = \left(\frac{p(\bsvarphi=\bsvarphi_0|s,\mathcal{H}_1)}{\pi(\bsvarphi=\bsvarphi_0|\mathcal{H}_1)}\right)^{-1}\label{eq:savagedickey}.
\end{align}

Note that for a signal $s$ with true parameters $\bstheta_s = (\bspsi_s,\bsvarphi_s)$ inferred assuming the subset hypothesis, the MAP estimate
\begin{align}
    \bstheta_{\rm MAP}|\mathcal{H}_0 = (\bspsi_{\rm MAP},\bsvarphi_0)|\mathcal{H}_1 \neq (\bspsi_s,\bsvarphi_s)|\mathcal{H}_1
\end{align}
can be significantly biased depending on the degree of correlations between $\bspsi$ and $\bsvarphi$, as previously shown for various GW sources~\cite{Kejriwal:2023djc,Garg:2024oeu,Chandramouli:2024vhw}. While the Bayes factor can be %conventionally 
calculated from Eq.~\eqref{eq:savagedickey} by sampling only from the $\mathcal{H}_1$ posterior, this can still be practically infeasible when multiple hypotheses are in contention. In the next section, we propose an alternate technique utilizing the importance sampling method, particularly suitable for a subclass of effects that perturbatively modify the subset hypothesis such that the induced biases can be linearly approximated.  This ``beyond-vacuum-GR'' class of effects includes tests of GR~\cite{Yunes:2009ke,Barausse:2016eii,LISA:2022kgy,Speri:2024qak}, astrophysical environments~\cite{Barausse:2014tra,Speri:2022upm,Kocsis:2011dr,Garg:2024qxq,Duque:2024mfw}, and even vacuum-GR consistent effects like mildly-eccentric binary evolution in stellar-mass black hole binaries~\cite{Favata:2013rwa,Romero-Shaw:2020thy,Romero-Shaw:2021ual}, secondary spin in EMRIs~\cite{Piovano:2020zin,Mathews:2021rod,Skoupy:2024uan}, etc., all of which perturbatively modify the subset hypothesis.

\begin{table}[]
    \centering
    \begin{tabular}{|c|c|}
        \hline
        \textbf{Symbol} & \textbf{Description} \\
        \hline
        \multirow{2}{*}{$\bspsi$} & parameters of the null\\
        &  (subset) model $\mathcal{H}_0$\\
        \hline
        \multirow{2}{*}{$\bsvarphi$} & extra parameters in\\
        &  the superset model $\mathcal{H}_1$\\
        \hline
        \multirow{2}{*}{$\bsvarphi_0$} & null value of $\bsvarphi$\\
        & which recovers $\mathcal{H}_0$\\
        \hline
    \end{tabular}
    \caption{Notation for the decomposition of the parameter vector $\bstheta$ in the nested modeling setup.}
    \label{tab:notation}
\end{table}

%%%%%%%%%%%%%%%%%%%%%%%%%%%%%%%%%%%%%%%%%%%%%
\section{Bias-Corrected Importance Sampling - General Formalism}\label{sec:methods}
%%%%%%%%%%%%%%%%%%%%%%%%%%%%%%%%%%%%%%%%%%%%%

%%%%%%%%%%%%%%%%%%%%%%%%%%%%%%%%%%%%%%%%%%%%%
\subsection{Constructing a bias-corrected proposal distribution}\label{sec:basicimplementation}
%%%%%%%%%%%%%%%%%%%%%%%%%%%%%%%%%%%%%%%%%%%%%

\begin{figure}
    \centering
    \begin{subfigure}{0.23\textwidth}
        \centering
        \includegraphics[trim=20 25 20 25,clip,width=\textwidth]{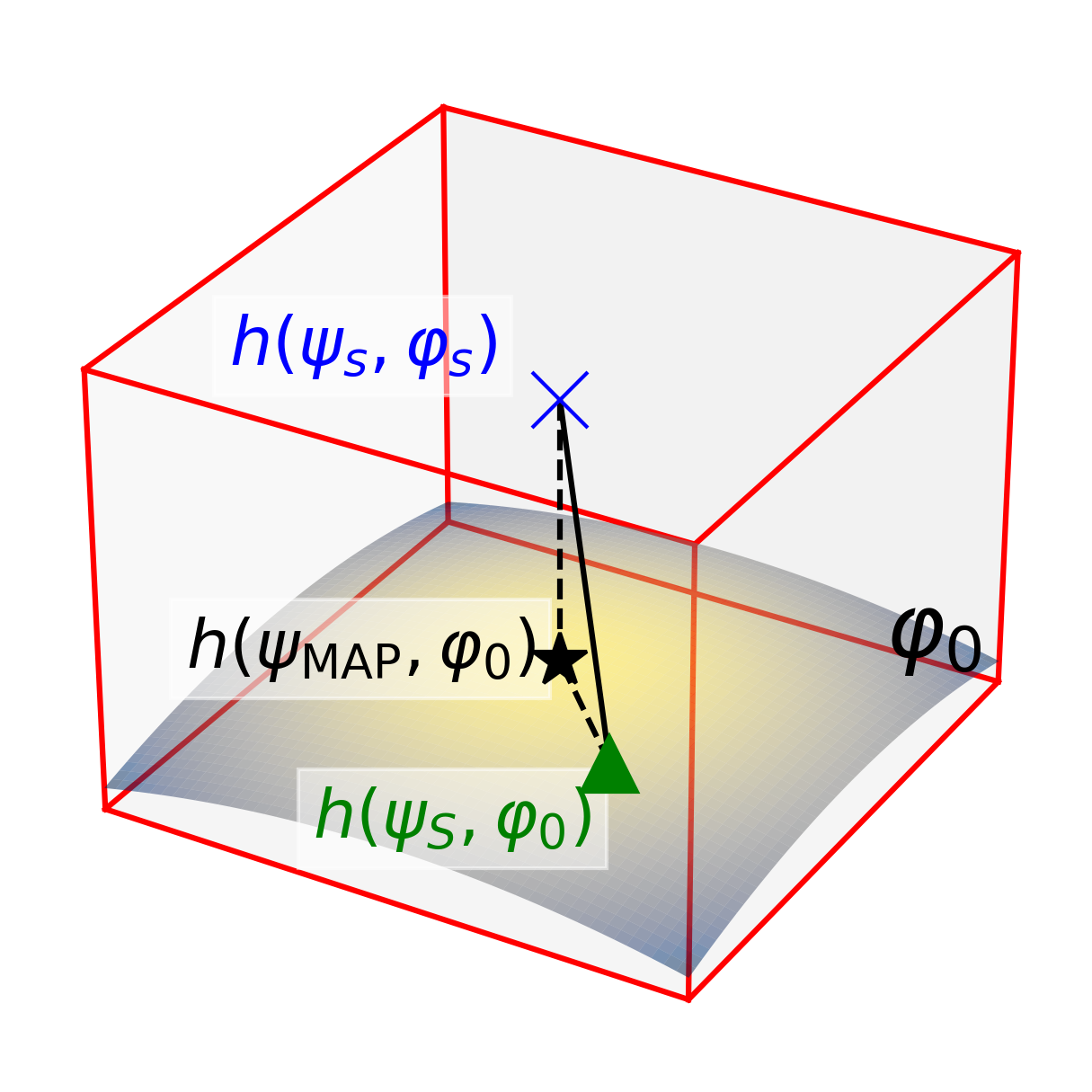}
    \end{subfigure}
    \begin{subfigure}{0.23\textwidth}
        \centering
        \includegraphics[trim=20 25 20 25,clip,width=\textwidth]{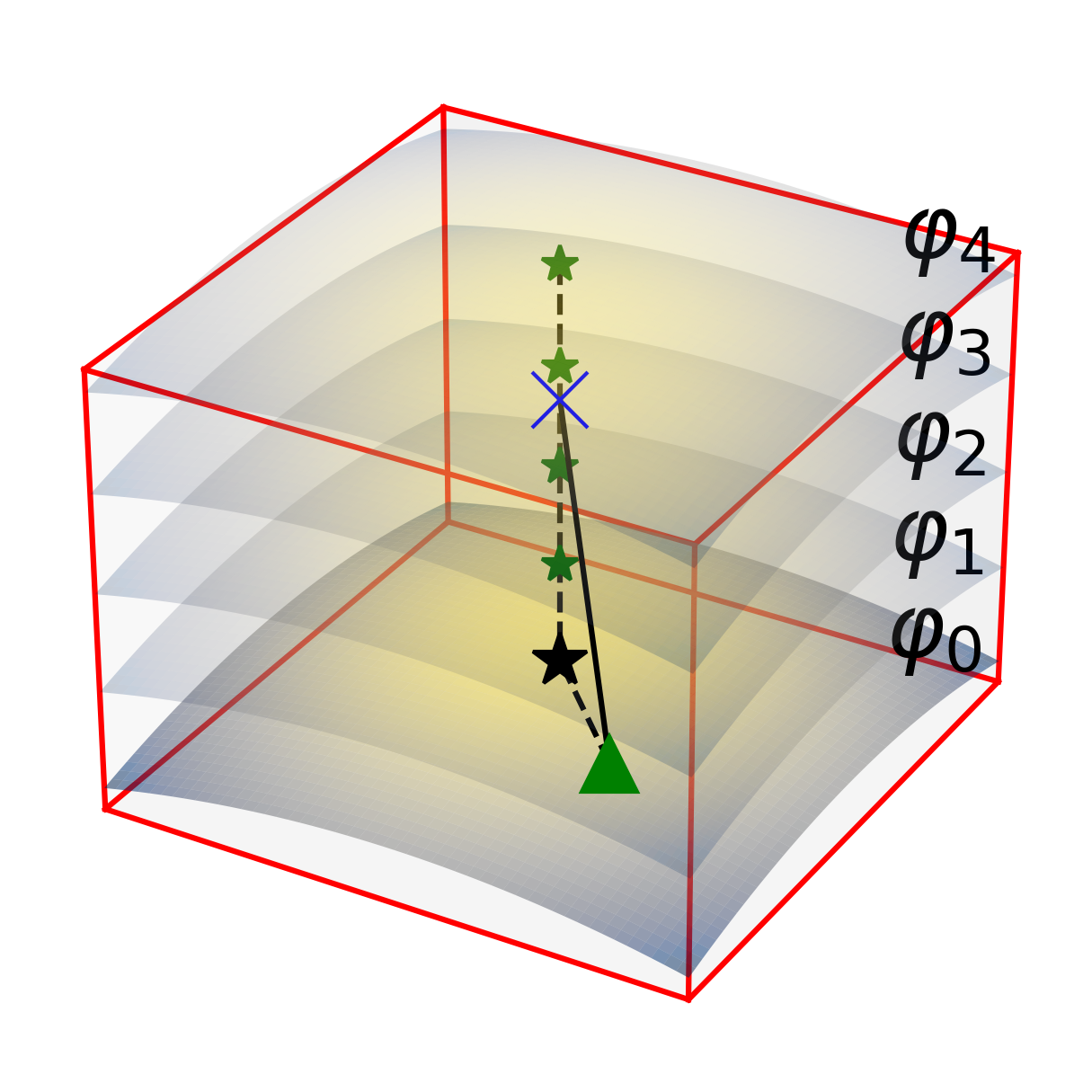}
    \end{subfigure}
    \caption{\justifying (Left panel) A schematic depicting an injected signal $h(\bspsi_s,\bsvarphi_s)$ (blue cross) generated in the superset hypothesis $\mathcal{H}_1$ (red hypercube) inferred assuming the subset hypothesis $\mathcal{H}_0$ (shaded submanifold) where $\bsvarphi = \bsvarphi_0$ is fixed. The green triangle is the signal's restriction to $\mathcal{H}_0$,  and the black star is the inferred MAP $h(\bspsi_{\rm MAP},\bsvarphi_0)$. The two points do not coincide, representing inference biases. (Right panel) 5 samples (green and black stars) drawn by translating the MAP along the correction axis (vertical dashed line) constructed following Eqs.~\eqref{eq:cvnested},~\eqref{eq:transformationfunction}.} 
    \label{fig:nestedvisualization}
\end{figure}

Our goal is to construct a proposal pdf $q(\bspsi,\bsvarphi)$ that roughly overlaps the true posterior $p(\bspsi,\bsvarphi|s,\mathcal{H}_1)$, i.e., the target distribution. We treat as given the set of samples $\{\bspsi\}|s,\mathcal{H}_0 \sim p(\bspsi|s,\mathcal{H}_0)$ of size $N$ in the subset hypothesis\footnote{The method of obtaining samples $\{\bspsi\}|\mathcal{H}_0$ is irrelevant as long as the samples are independent and identically drawn from $p(\bspsi|s,\mathcal{H}_0)$.}, with $\bspsi_{\rm MAP}|\mathcal{H}_0$ as the MAP\footnote{In the following, we denote $\bspsi_{\rm MAP} \equiv \bspsi_{\rm MAP}|\mathcal{H}_0$ for brevity unless stated otherwise.}. For a perturbative effect parameter vector $\bsvarphi_s$ in the true signal $s$ and assuming diffuse priors, the induced bias $\Delta\bspsi := \bspsi_{\rm MAP} - \bspsi_s$ on the set of common parameters can be linearly approximated as~\cite{Kejriwal:2023djc}
\begin{align}
    \Delta\bspsi = (\Gamma_{\bspsi\bspsi})^{-1}\Gamma_{\bspsi\bsvarphi}\cdot(\bsvarphi_s - \bsvarphi_0),\label{eq:cvnested}
\end{align}
where $\Gamma$ is the GW Fisher information matrix (FIM) with elements~\cite{Cutler:1994ys,Vallisneri:2007ev}
\begin{align}
    \Gamma_{ij} := \left<\left.\frac{\partial h}{\partial \bstheta_i}\right|\frac{\partial h}{\partial \bstheta_j}\right> \, ,
\end{align}
evaluated at $(\bspsi_{\rm MAP},\bsvarphi_0)$ in the superset hypothesis $\mathcal{H}_1$. The left panel of Fig.~\ref{fig:nestedvisualization} visualizes the biased inference, where the MAP is obtained at the projection of $h(\bspsi_s,\bsvarphi_s)$ onto the subset manifold (i.e., where $\bsvarphi=\bsvarphi_0$ is fixed). Eq.~\eqref{eq:cvnested} incorporates biases induced by parameter correlations (through $\Gamma_{\bspsi\bsvarphi}$) and the strength of the missing effect (through $\bsvarphi_s - \bsvarphi_0$). If the true value of $\bsvarphi_s$ is known, the parameters of the signal can be recovered approximately as $(\bspsi_s,\bsvarphi_s) = \tau(\bspsi_{\rm MAP},\bsvarphi_s)$ where
\begin{align}
    \tau(\bspsi,\bsvarphi) := (\bspsi - \mathcal{T}\cdot(\bsvarphi - \bsvarphi_0), \, \mathcal{I}(\bsvarphi))\label{eq:transformationfunction} \, , 
\end{align}
is a function of $\bspsi$ and $\bsvarphi$ with codomain $\mathbb{R}^{d_{\bspsi}}\times\mathbb{R}^{d_{\bsvarphi}}$, $\mathcal{I}$ is the identity function, and we define the shorthand $\mathcal{T}:= (\Gamma_{\bspsi\bspsi})^{-1}\Gamma_{\bspsi\bsvarphi}$ for the translation factor.

Without access to $\bsvarphi_s$, we instead draw a set of $m$ samples $\bsvarphi|\mathcal{H}_1 \sim \pi(\bsvarphi|\mathcal{H}_1)$ from a suitable prior distribution on $\bsvarphi$, and form the Cartesian product $\{(\bspsi,\bsvarphi)\}:=\{\bspsi|s,\mathcal{H}_0\}\times\{\bsvarphi|\mathcal{H}_1\}$ of size $N\times m$. This set of samples is then transformed via Eq.~\eqref{eq:transformationfunction} as $\{(\bspsi',\bsvarphi'):=\tau(\bspsi,\bsvarphi)\}$; the transformed samples are distributed according to a proposal pdf $q(\bspsi',\bsvarphi')$, whose explicit form we derive below. First, note that a sample $\bspsi|s,\mathcal{H}_0 \sim p(\bspsi|s,\mathcal{H}_0)$ being paired with an independent draw $\bsvarphi|\mathcal{H}_1 \sim \pi(\bsvarphi|\mathcal{H}_1)$ has the joint density
\begin{align}
    g(\bspsi,\bsvarphi) = p(\bspsi|s,\mathcal{H}_0)\pi(\bsvarphi|\mathcal{H}_1) .
\end{align}
The proposal pdf is similarly the joint density of obtaining $(\bspsi',\bsvarphi') = \tau(\bspsi,\bsvarphi)$,
whose measure can be expressed as the push-forward of $g(\bspsi,\bsvarphi)$,
\begin{align*}
    q(\bspsi',\bsvarphi') &:= g'(\bspsi',\bsvarphi') \\
    &= g(\tau^{-1}(\bspsi',\bsvarphi'))\cdot |J|\\
    &= g(\bspsi'+\mathcal{T}\cdot(\bsvarphi'-\bsvarphi_0),\bsvarphi')\cdot|J|\numberthis
\end{align*}
where $J$ is the Jacobian matrix of the inverse transformation, with determinant
\begin{align}
    |J| = \begin{vmatrix}
        \frac{\partial \bspsi}{\partial\bspsi'} & \frac{\partial \bspsi}{\partial\bsvarphi'}\\
        \\
        \frac{\partial \bsvarphi}{\partial\bspsi'} & \frac{\partial \bsvarphi}{\partial\bsvarphi'}\\
    \end{vmatrix} = \begin{vmatrix}
        I_{d_{\bspsi}} & \mathcal{T}\\
        0 & I_{d_{\bsvarphi}}\\
    \end{vmatrix} = 1.
\end{align}
for a $d\times d$ identity matrix $I_{d}$. Thus, 
\begin{align*}
    q(\bspsi',\bsvarphi') &= g(\bspsi,\bsvarphi)\\
    & = p(\bspsi|s,\mathcal{H}_0)\pi(\bsvarphi|\mathcal{H}_1)\\
    &\propto \mathcal{L}(\bspsi|s,\mathcal{H}_0)\pi(\bspsi|\mathcal{H}_0)\pi(\bsvarphi|\mathcal{H}_1).\numberthis\label{eq:basicbiascorrectedproposal}
\end{align*}
The right panel in Fig.~\ref{fig:nestedvisualization} schematizes the constructed bias-corrected proposal pdf. The shaded 2D sub-manifolds represent slices of fixed $\bsvarphi$ but variable $\bspsi$, and the $\bsvarphi_0$ manifold denotes the subset hypothesis, $\mathcal{H}_0$. The red hypercube represents the space of variable $\bsvarphi$, i.e. the superset hypothesis, $\mathcal{H}_1$. The green stars denote a set of proposed samples obtained by translating the MAP along the ``correction axis'' (dashed vertical line) informed by $\bsvarphi$'s correlations with $\bspsi$ through Eqs.~\eqref{eq:cvnested},~\eqref{eq:transformationfunction}.

\subsection{Importance sampling formalism for the bias-corrected proposal}
Given the set of samples $\{\bspsi',\bsvarphi'\}
\sim q(\bspsi',\bsvarphi')$,
the (unnormalized) importance weights are
\begin{align}
    {w}_{ij} &:= \frac{p(\bspsi_i',\bsvarphi_j'|s,\mathcal{H}_1)}{q(\bspsi_i',\bsvarphi_j')}.~\label{eq:importanceweightsbasic}
\end{align}
Here, $p(\bspsi',\bsvarphi'|s,\mathcal{H}_1)$ is the target pdf, given by Bayes' theorem as
\begin{align}
    p(\bspsi',\bsvarphi'|s,\mathcal{H}_1) = \frac{\mathcal{L}(\bspsi',\bsvarphi'|s,\mathcal{H}_1)\pi(\bspsi',\bsvarphi'|\mathcal{H}_1)}{\mathcal{Z}(s|\mathcal{H}_1)}.\label{eq:targetpdf}
\end{align} 
Following a similar push-forward
argument as Eq.~\eqref{eq:basicbiascorrectedproposal}, we have \begin{align*}
    \pi(\bspsi',\bsvarphi'|\mathcal{H}_1) &= \pi(\bspsi,\bsvarphi|\mathcal{H}_1)\\
    &=\pi(\bspsi|\mathcal{H}_0)\pi(\bsvarphi|\mathcal{H}_1).\numberthis\label{eq:priorpdftransform}
\end{align*} 
Using Eqs.~\eqref{eq:basicbiascorrectedproposal},~\eqref{eq:targetpdf}, and~\eqref{eq:priorpdftransform}, the importance weights (Eq.~\eqref{eq:importanceweightsbasic}) can thus be rewritten as
\begin{align}
    w_{ij} = \frac{\mathcal{L}(\bspsi'_i,\bsvarphi'_j|s,\mathcal{H}_1)}{\mathcal{L}(\bspsi_i|s,\mathcal{H}_0)}.\label{eq:likeweights}
\end{align}
Given the prior probability of obtaining $\bsvarphi = \bsvarphi_0$ in $\mathcal{H}_1$, $\pi(\bsvarphi=\bsvarphi_0|\mathcal{H}_1)$, and the posterior probability $p(\bsvarphi=\bsvarphi_0|s,\mathcal{H}_1)$ from Eqs.~\eqref{eq:importanceweightsbasic} and~\eqref{eq:likeweights}, the Savage-Dickey ratio can be straightforwardly evaluated following Eq.~\eqref{eq:savagedickey}.\footnote{In sampling-based inference, it is standard to bin values around $\bsvarphi_0$ to approximate $p(\bsvarphi=\bsvarphi_0) \approx p(\Vert\bsvarphi-\bsvarphi_0\Vert<b)$ for some bin size $b$ and where $\Vert\cdot\Vert$ denotes the norm.}

%%%%%%%%%%%%%%%%%%%%%%%%%%%%%%%%%%%%%%%%%%%%%
\subsection{A more conservative proposal pdf based on a regularized normal approximation}\label{sec:regularizedimplementation}
%%%%%%%%%%%%%%%%%%%%%%%%%%%%%%%%%%%%%%%%%%%%%
The proposal constructed above aims to correct the induced inference biases through $\tau(\bspsi,\bsvarphi)$ to maximize the overlap with the target pdf. However, the method strongly relies on the ability to accurately estimate the FIM at $\{\bspsi_{\rm MAP},\bsvarphi_0\}$ in $\mathcal{H}_1$, which can be difficult if, e.g., the model parametrization is nearly ``ill-posed'', such that the FIM is poorly-conditioned, increasing its susceptibility to numerical errors~\cite{Vallisneri:2007ev}. Here, we attempt to mitigate some of these technical challenges by constructing a more conservative proposal pdf, as described below.

We again treat as given the set of samples $\{\bspsi\}|\mathcal{H}_0 \sim p(\bspsi|s,\mathcal{H}_0)$. However, instead of randomly drawing $\bsvarphi\sim \pi(\bsvarphi|\mathcal{H}_1)$, we deterministically choose $\bsvarphi = (\bsvarphi_1, \ldots, \bsvarphi_j,\ldots,\bsvarphi_m)$ from a regular grid of size $m$ bounded by some range $[\bsvarphi_{\rm min},\bsvarphi_{\rm max}]$. The proposal is constructed in $m$ iterations, where at the $j^{\rm th}$ step, we first obtain the transformed set of samples $\bstheta'_j = \tau(\bspsi,\bsvarphi_j)$ and calculate their posterior expectation $\bar{\bstheta}'_j := \mathbb{E}[\bstheta_j']$ by averaging. Then, $N$ samples of $(\bspsi'_{j},\bsvarphi'_j)$ are redrawn from a \textit{regularized normal approximation} of the likelihood at $\bar{\bstheta}_j'$,
\begin{align}
    q_j(\bspsi',\bsvarphi') := \frac{|\Sigma_{\rm reg}|^{-1/2}}{(2\pi)^{d/2}}\exp[-\frac{1}{2}(\bstheta'-\bar\bstheta'_j)^T\Sigma_{\rm reg}^{-1}(\bstheta'-\bar\bstheta'_j)]
\end{align}
where
\begin{align}
    \Sigma_{\rm reg} := \Gamma^{-1} + \epsilon\cdot D \label{eq:regularize}
\end{align}
is the regularized inverse of the FIM~\cite{tikhonov1977solutions,481c956f-fc76-3a47-912c-132550d4ccb6} with $\epsilon > 0$ some scalar regularization factor and $D$ a $d\times d$ diagonal matrix. We thus form a set of $N\times m$ samples $\{\bspsi',\bsvarphi'\}:= \bigcup_j \{\bspsi'_{j},\bsvarphi'_j\}$, which is distributed according to the mixture of all the regularized normal elements, \begin{align}
    q_{\rm reg}(\bspsi',\bsvarphi') := \frac{1}{m}\sum_j q_j(\bspsi',\bsvarphi').
\end{align}
Here, the choice of $m$ is left as a free parameter of the method for a fair comparison with the basic implementation in the examples presented in the next section. It may also be informed, e.g., by the fractional coverage of the prior box by $\Sigma_{\rm reg}$ to ascertain sufficient overlap between consecutive $q_j$.

This choice of the proposal distribution assumes that the FIMs evaluated at the true signal point $(\bspsi_s,\bsvarphi_s)|\mathcal{H}_1$ and the vacuum MAP point $(\bspsi_{\rm MAP},\bsvarphi_0)|\mathcal{H}_1$ are equal, which is consistent with the linear signal approximation~\cite{Cutler:2007mi,Kejriwal:2023djc}. We center the $j^{\rm th}$ proposal element $q_j$ at $\bar\bstheta_j'$ instead of the translated MAP ($\tau(\bspsi_{\rm MAP},\bsvarphi_j)$) since the sample mean is a more robust estimate of the MAP for the symmetric posteriors expected for high-SNR sources, and since the linear signal approximation fails for low-SNR asymmetric pdfs anyway. Furthermore, since $q_j$ is normally distributed, which is symmetric, it is reasonable to centre it at the mean of the translated samples. With the additional regularization step, we effectively introduce a noisy baseline to the inverse FIM elements to account for numerical uncertainties in its calculation and inversion.~\footnote{Empirically, we found regularizing $\Gamma^{-1}$ to yield better results compared to directly regularizing $\Gamma$ in all examples presented in the next section.} We choose
\begin{align}
    \epsilon &= \frac{|\Gamma\cdot\Gamma^{-1}|-1}{d}\label{eq:regularizationfactor}\\
    D &= {\rm diag}(\Gamma^{-1}) \label{eq:regularizationmatrix}
\end{align}
such that the noisy baseline scales with the inversion errors through $\epsilon$ and the uncertainties in the model parameters through $D$. The regularization step in Eq.~\eqref{eq:regularize} can be implemented iteratively, such that at the $k^{\rm th}$ iteration, 
\begin{align}
    \Sigma_{{\rm reg},k} &= \Gamma_{k-1}^{-1} + \epsilon_k \cdot D_k
\end{align}
where $\Gamma_{k} := \Sigma_{{\rm reg},k}^{-1}$, $\Gamma_{0} \equiv \Gamma$, and
\begin{align}
    \epsilon_k &:= \frac{|\Gamma_{k-1}\cdot \Gamma_{k-1}^{-1}|-1}{d}\label{eq:iterativeepsilon}\\
    D_k &:= {\rm diag}(\Gamma_{k-1}^{-1})\label{eq:iterativeD}
\end{align}
until the condition $\epsilon_k<\epsilon_0$ is met, where $\epsilon_0>0$ is a suitable error tolerance. 

\begin{figure}
    \centering
    \includegraphics[width=0.95\linewidth]{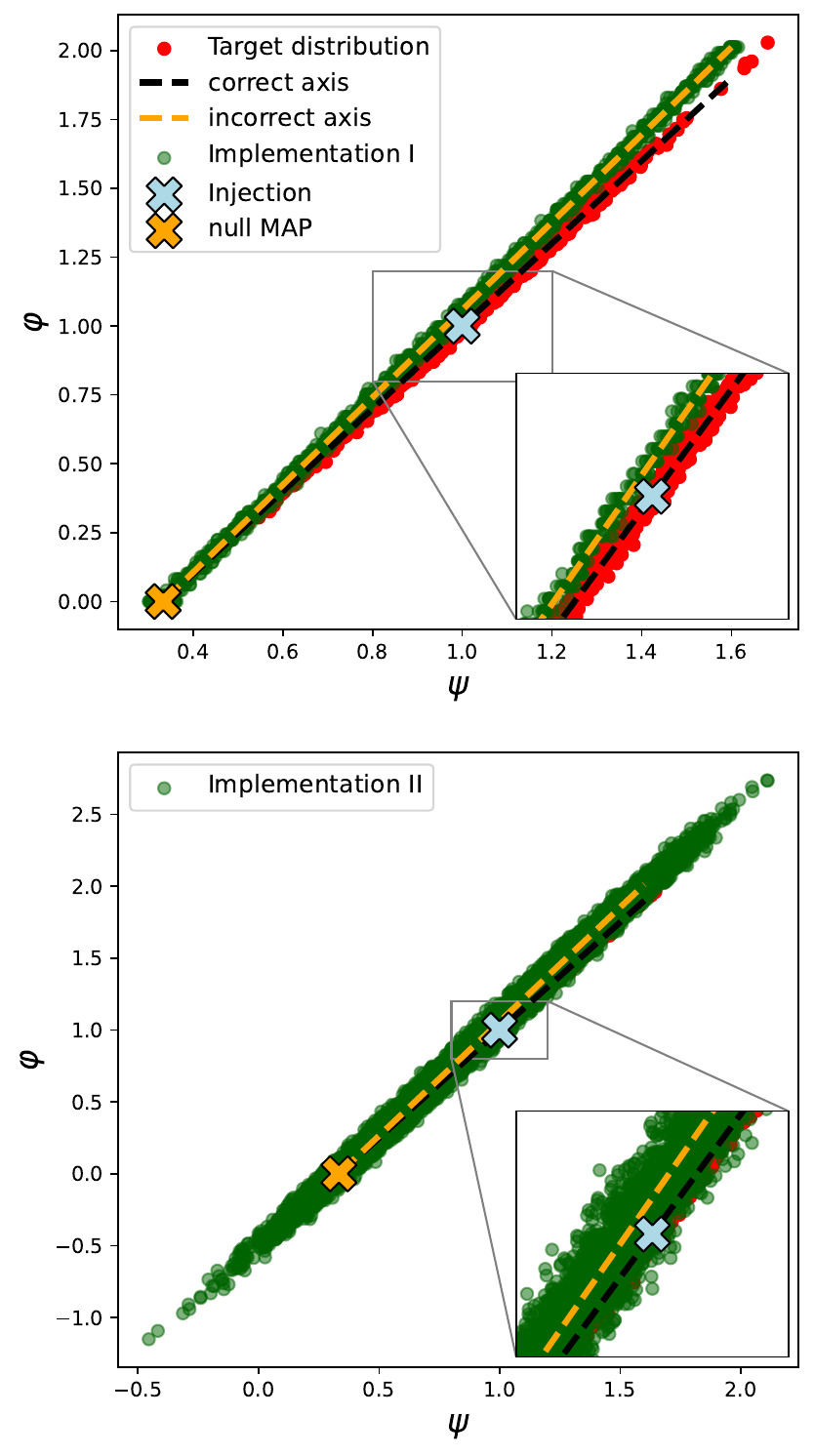}
    \caption{\justifying A 2D schematic with the subset hypothesis parameter $\psi$ along the x-axis and the superset hypothesis parameter $\varphi$ along the y-axis. The injection (blue cross) is recovered at the MAP (orange cross) with $\bsvarphi=0.0$ in the null hypothesis. We depict a case where the constructed correction axis (orange dashed line) deviates from the truth (black dashed line) due to computational errors. Draws from the target distribution are depicted in red and are overlayed by draws from the basic implementation (Sec.~\ref{sec:basicimplementation}, top panel) and the regularized implementation (Sec.~\ref{sec:regularizedimplementation}, bottom panel). The conservative proposals in the second method enable better coverage of the target, even when the correction axis is erroneous.}
    \label{fig:regularizedschematic}
\end{figure}

We present a schematic visualization of the practical implications of the regularized proposal in Figure~\ref{fig:regularizedschematic}, using only two parameters: the null hypothesis parameter $\psi$ along the x-axis and the additional superset hypothesis parameter $\varphi$ along the y-axis. The calculated correction axis (orange dashed line) is shown to deviate slightly from the true correction axis $(\psi_s-\psi_\mathrm{MAP},\varphi_s-\varphi_0)$ (black dashed line) --- this however is enough to inhibit sufficient overlap between the constructed proposals (green) and the target pdf (red) in the basic implementation, as shown in the top panel. In the regularized implementation with $\epsilon = 10^{-2}$ and $D$ given by Eq.~\eqref{eq:regularizationmatrix}, the coverage of the same target pdf improves significantly as presented in the bottom panel. By construction, however, the regularized proposals also cover a larger ``empty space'' around the target distribution compared to the basic implementation, which may reduce the method's importance efficiency $\eta$. This is consistent with the example results quoted in the next section, where the importance efficiency of the basic implementation was found to be factor $\sim 2-3$ higher than in the regularized framework.

Given the set of proposed samples $(\bspsi',\bsvarphi')\sim q_{\rm reg}(\bspsi',\bsvarphi')$, the (unnormalized) importance weights are given following arguments from the previous section as 
\begin{align}
    w_{{\rm reg},i} &=\frac{p(\bspsi'_i,\bsvarphi'_i|s,\mathcal{H}_1)}{q_{\rm reg}(\bspsi'_i,\bsvarphi'_i)}.
\end{align}
Again, given the prior probability $\pi(\bsvarphi=\bsvarphi_0|\mathcal{H}_1)$, and the posterior distribution $p(\bsvarphi=\bsvarphi_0|\mathcal{H}_1)$ as the marginalization of $p(\bspsi',\bsvarphi'|s,\mathcal{H}_1)$ over $\bspsi'$ evaluated at $\bsvarphi_0$, we obtain the Savage-Dickey ratio following Eq.~\eqref{eq:savagedickey}.

%%%%%%%%%%%%%%%%%%%%%%%%%%%%%%%%%%%%%%%%%%%%%
\section{Examples}\label{sec:examples}
%%%%%%%%%%%%%%%%%%%%%%%%%%%%%%%%%%%%%%%%%%%%%

\subsection{Setup}

\noindent \textit{Waveform model} --- We now present example results contextualizing the two bias-corrected techniques---the \textit{basic} and \textit{regularized} implementations---for LISA EMRIs. The injected signal $s$ is modeled as a fully relativistic adiabatic Kerr EMRI~\cite{Pound:2021qin} evolving in a circular-equatorial orbit, implemented by Khalvati et al.~\cite{Khalvati:2024tzz} in the modular framework of the \texttt{FastEMRIWaveforms (FEW)} package~\cite{Katz:2021yft,Chua:2020stf}. It is described by 10 vacuum-GR parameters: the redshifted masses, $M$, $\mu$, of the MBH and CO respectively, the dimensionless spin of the MBH, $a$, the initial semi-latus rectum of the orbit, $p_0$, the initial azimuthal phase, $\Phi_0$, and extrinsic parameters $D_L, (\theta_S,\phi_S),(\theta_K,\phi_K)$, i.e. the source's luminosity distance, sky location, and the MBH's spin orientation to the solar system, respectively. The true source additionally incorporates the planetary migration effect due to an accretion disk surrounding the MBH, which can be modeled as a secular power-law perturbation modifying the flux balance laws of the EMRI evolution as~\cite{Kocsis:2011dr,Barausse:2014tra,Speri:2022upm}
\begin{align}
    \dot{L} = \dot{L}_{\rm GR}\left(1 + A\left(\frac{p}{10M}\right)^{n_r}\right)
\end{align}
where $\dot{L}_{\rm GR}$ is the axial component of the angular momentum flux in vacuum-GR calculated at leading order, $p$ is the semi-latus rectum of the instantaneous orbit, and $(A, n_r)$ are 2 new parameters that describe the disk effect. In the following examples, the accretion + vacuum-GR evolution with variable $A$ (and some fixed $n_r$) characterizes the superset beyond-vacuum-GR hypothesis ($\mathcal{H}_1$), and $A=0$ is fixed in the subset vacuum-GR hypothesis ($\mathcal{H}_0$).

\noindent \textit{Data analysis setup} --- We asuume flat priors on all model parameters such that the posterior is proportional to the likelihood and the MAP is equal to the maximum likelihood estimate (MLE). We sample from the various posteriors in our study using \texttt{Eryn}, a Bayesian inference package designed for LISA~\cite{Karnesis:2023ras}. It incorporates parallel tempering useful for inferring biased MAPs, and interfaces with \texttt{LISAanalysistools (LAT)}~\cite{michael_katz_2024_10930980} and \texttt{fastlisaresponse}~\cite{Katz:2022yqe} utility packages, providing consistent and modular integration with GW data analysis tools and LISA noise models. To accurately calculate the FIM at the MLE, required for the transformation function, we employ the \texttt{StableEMRIFishers (SEF)}~\cite{kejriwal_2024_sef} package that systematically calculates numerical waveform derivatives and hence the FIM, enhancing its stability.

In all the following examples, we work in the log-mass parametrization $(M,\mu)\to (\ln M, \ln \mu)$, which reduces the ill-posedness of the model, and choose $M = 10^6 M_\odot$, $\mu = 50 M_\odot$, $a = 0.9$, $\Phi_0 = 1.0$, $(\theta_S,\phi_S) = (\pi/4,\pi/3)$, and $(\theta_K,\phi_K)= (\pi/6,\pi/5)$ in the true signal. We additionally fix $n_r = 8$ corresponding to the inner region of a typical AGN disk~\cite{Kocsis:2011dr, Speri:2022upm}. $p_0$ is chosen for the EMRI to plunge at the end of the 1-year observation window, and $D_L$ such that the source's signal-to-noise ratio (SNR) $\rho := \langle s|s\rangle^{1/2} = 100  $. Bayesian inference is performed with 8 MCMC walkers across 3 temperatures within a reasonable prior hypercube (flat, bounded priors) centered at the injected signal parameters. Finally, given the injected parameter vector $\hat{\bspsi}$, the mean of the MCMC samples from the subset hypothesis, $\bar\bspsi_{{\rm sub}}$, and the corresponding mean of the importance samples in the superset hypothesis, $\bar\bspsi_{{\rm sup}}$ (Eqs.~\eqref{eq:numericalexpect} and~\eqref{eq:likeweights}), we define
\begin{align}
    r_{i} := \frac{|\hat{\bspsi}_i-\bar{\bspsi}_{{\rm sub},i}|}{|\hat{\bspsi}_i-\bar{\bspsi}_{{\rm sup},i}|}\label{eq:fractionalimprovement}
\end{align}
as a measure of the fractional change in the point estimate of $\bspsi$. $r_i > 1$ corresponds to an improved (closer to the injection) expectation value of the $i^{\rm th}$ model parameter after the procedure.

\subsection{Example \rom{1}: Intrinsic parameters only}

\begin{figure}
    \centering
    \includegraphics[width=0.95\linewidth]{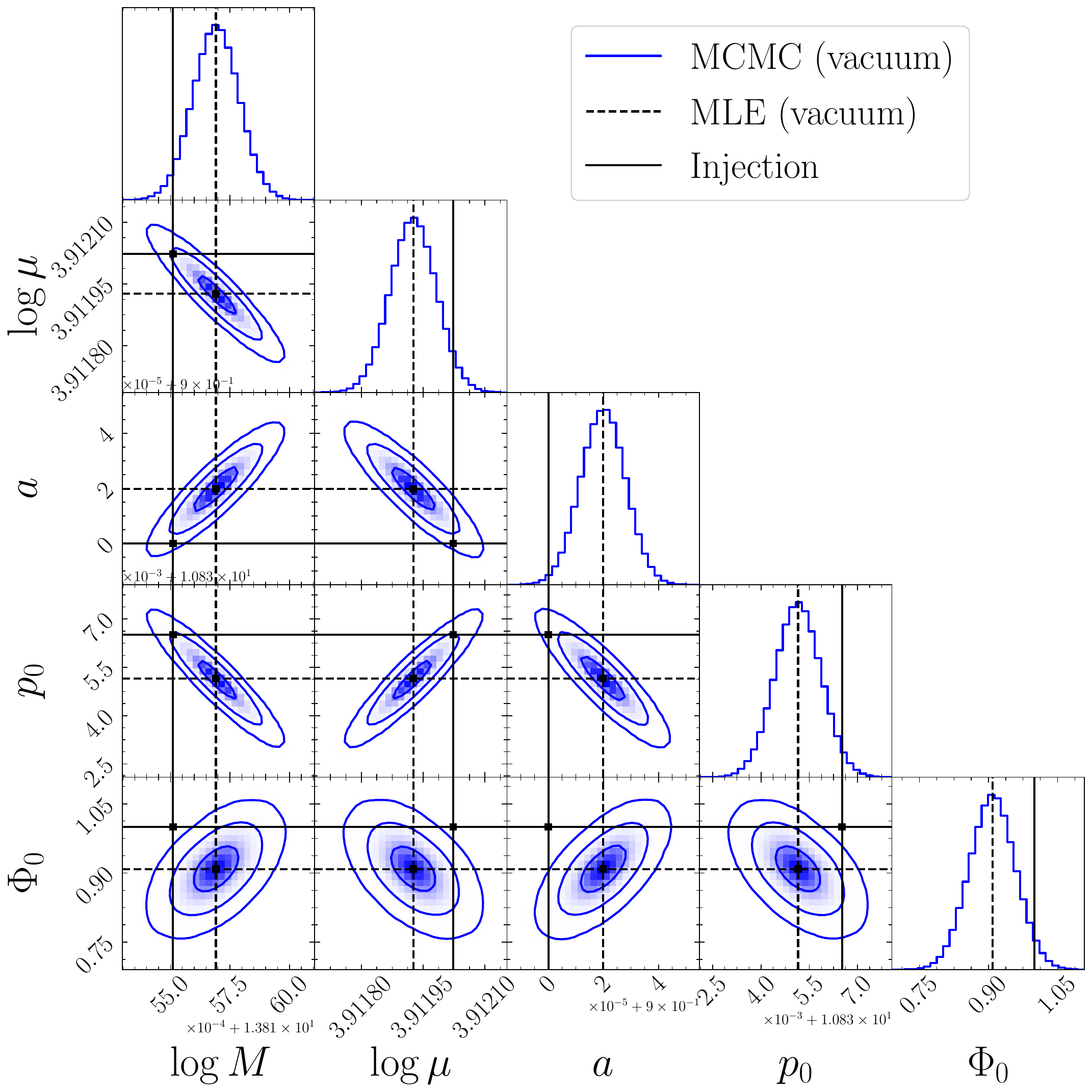}
    \caption{\justifying Distribution of posterior samples of an EMRI signal assuming the vacuum-GR (subset) hypothesis $\mathcal{H}_0$ with model parameters $\bspsi = (\ln M,\ln \mu,a,p_0,\Phi_0)$ (blue contours). The injected signal exhibits an additional beyond-vacuum-GR parameter $\bsvarphi = A = 1.92\times 10^{-5}$ induced by the effect of an accretion disk around the MBH. The recovered MLE $\bspsi_{\rm MLE}|\mathcal{H}_0$ (dashed line) is $\sim 2\sigma$ biased to the injected parameters (solid line).}
    \label{fig:5paramMCMC}
\end{figure}

\begin{figure}
    \centering
    \begin{subfigure}{0.45\textwidth}
        \centering
        \includegraphics[width=\textwidth]{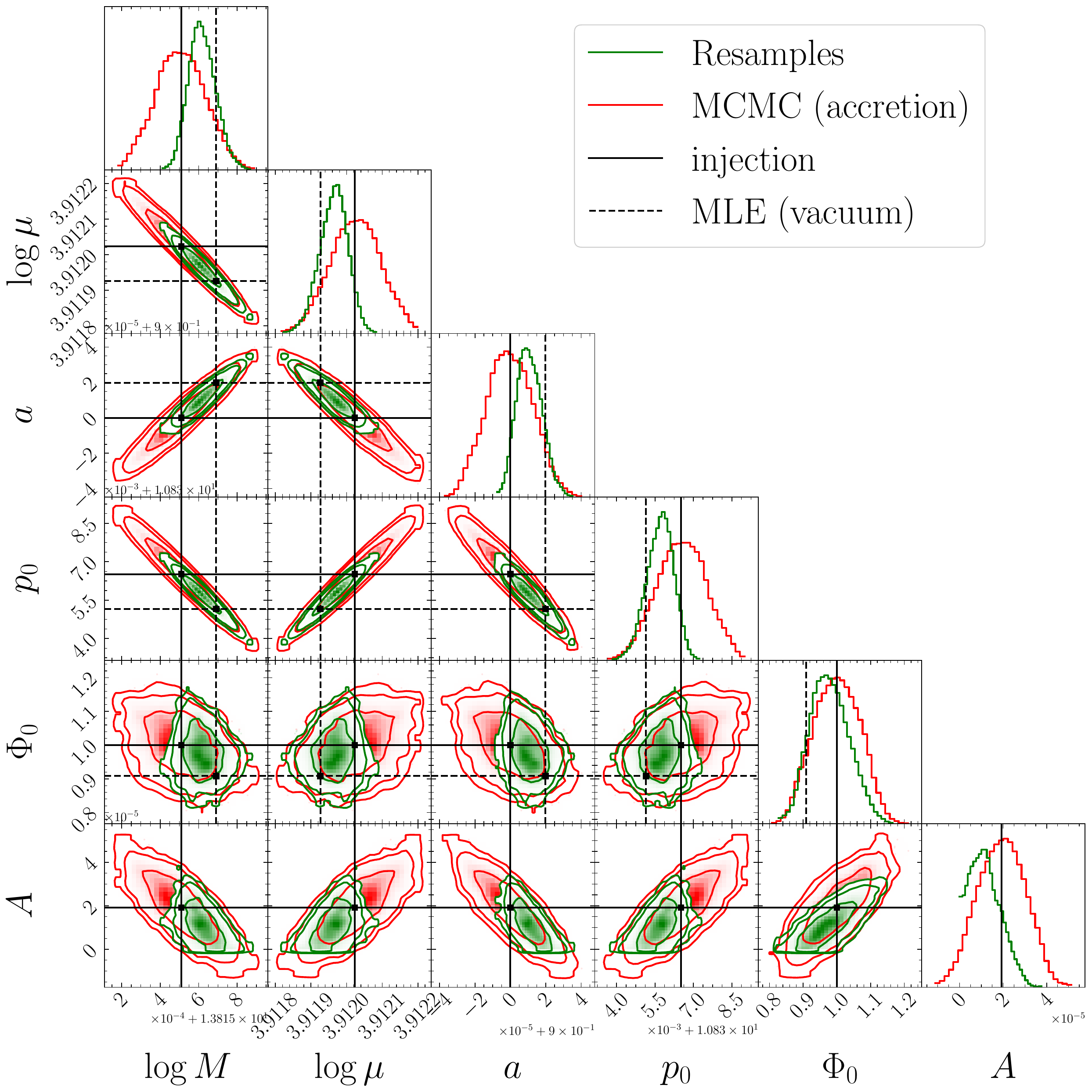}
    \end{subfigure}
    \hfill
    \hfill
    \begin{subfigure}{0.45\textwidth}
        \centering
        \includegraphics[width=\textwidth]{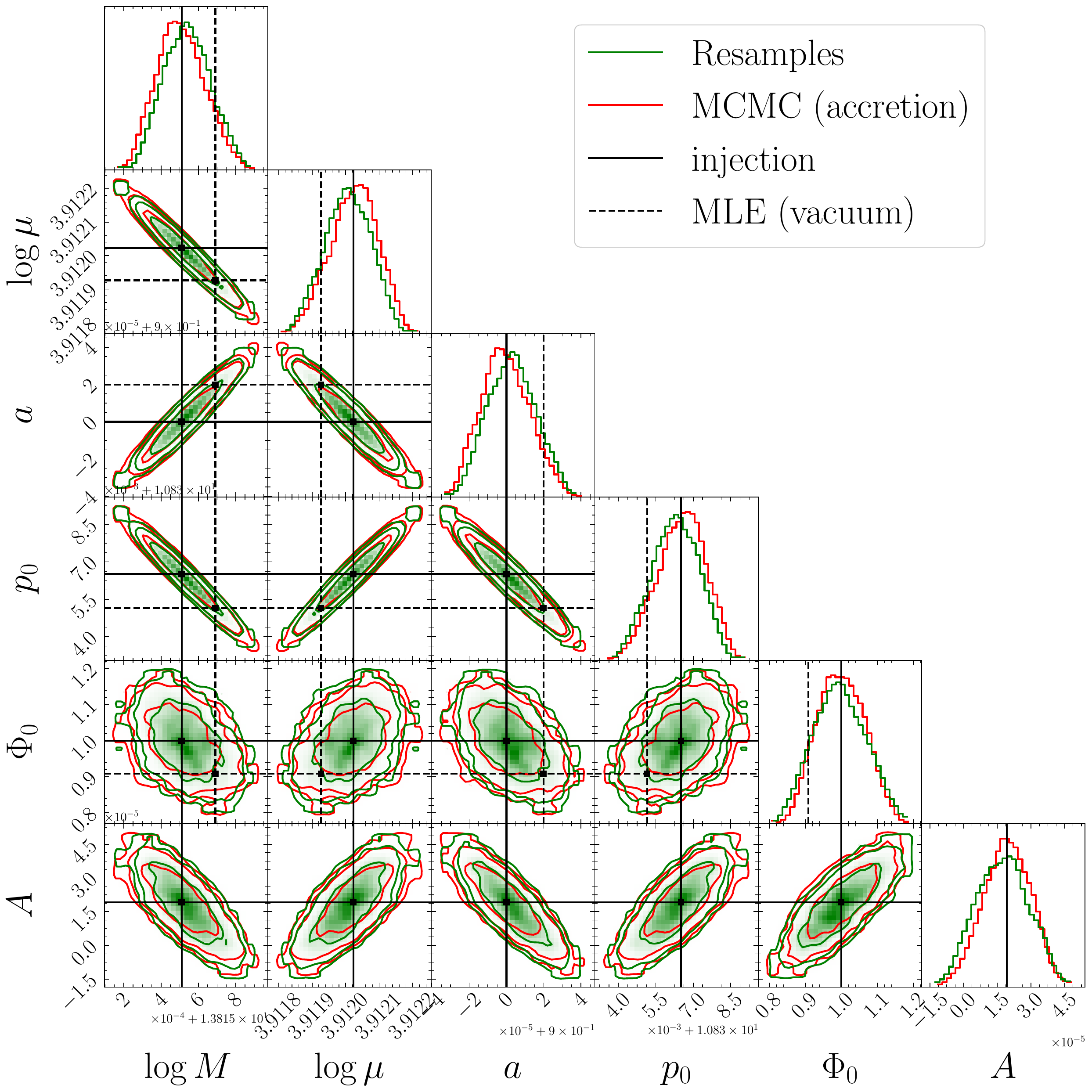}
    \end{subfigure}
    \caption{\justifying Distribution of bias-corrected importance \textit{re}samples (green contours) in Example \rom{1} in the basic implementation (section~\ref{sec:basicimplementation}, top panel) and regularized implementation (section~\ref{sec:regularizedimplementation}, bottom panel), overlayed with posterior samples of the injected EMRI signal in the true beyond-vacuum-GR (superset) hypothesis $\mathcal{H}_1$ (red contours).
    }
    \label{fig:5paramresamples}
\end{figure}

In the first example, all extrinsic parameters are fixed, i.e., only the parameter set $\bspsi := (\ln M,\ln \mu,a,p_0,\Phi_0)$ is inferred in the subset hypothesis. The superset hypothesis includes the additional parameter $\bsvarphi := (A)$, with injection value $A = 1.92 \times 10^{-5}$. In the subset hypothesis, we first generate $\approx 9\times10^5$ total MCMC samples $\bspsi|\mathcal{H}_0 \sim \mathcal{L}(\bspsi|s,\mathcal{H}_0)$ and found the MLE  $\bspsi_{\rm MLE}|\mathcal{H}_0$ at $\sim 2\sigma$ bias to the injection. As shown in the corner plot in Fig.~\ref{fig:5paramMCMC}, the distribution of these samples poorly overlaps with the injected signal parameters, especially along $a$ and $\Phi_0$, rendering standard importance sampling (i.e., with $\tau = I$) infeasible, and motivating the bias-correction procedure outlined above. We set the prior range on $A$ as $[A_{\rm min},A_{\rm max}] = [0.0,1.0]\times10^{-4}$, the size $N$ of samples in the subset hypothesis equal to the final $1\%$ of the MCMC samples, and fix $m=5$, for a total of $N\times m \approx 5\times 10^4$ samples of $(\bspsi,\bsvarphi)$. The number of discarded samples and choice of $m$ were empirically made to ensure a sufficient (but not excessive, for computational feasibility) set of proposal samples. Furthermore, we approximate the vacuum-GR MLE as $\bspsi_{\rm MLE} \approx \mathbbm{E}[\bspsi_{<3\sigma}|\mathcal{H}_0]$ where $\bspsi_{<3\sigma}|\mathcal{H}_0$ is the set of vacuum-GR MCMC samples within $3\sigma$ of the MLE, which is correspondingly the set of $99.7\%$ of samples with the highest likelihood values. The mean of a finite set of samples provides a more robust estimate of the MLE than the inferred MLE point for symmetric likelihood surfaces, as expected for high-SNR sources as in this example (see Fig.~\ref{fig:5paramMCMC}). Additionally, by only considering samples within $3\sigma$ of the MLE point, we ensure that any tail asymmetries in the sampling do not bias the estimate. Finally, the FIM is calculated at $(\bspsi_{\rm MLE},\bsvarphi_0)|\mathcal{H}_1$.

In the basic implementation, samples of $\bsvarphi \equiv A$ are drawn uniformly from its prior range and pairs of $(\bspsi,\bsvarphi)$ are transformed using $\tau$ (Eq.~\eqref{eq:transformationfunction}) to obtain the bias-corrected samples $\{\bspsi',\bsvarphi'\}\sim q(\bspsi',\bsvarphi')$, with the proposal distribution $q(\bspsi',\bsvarphi')$ given by Eq.~\eqref{eq:basicbiascorrectedproposal} (with $\pi(\bspsi|\mathcal{H}_0) \propto 1$). The importance resampled means of the vacuum-GR parameters are calculated following Eqs.~\eqref{eq:numericalexpect} and~\eqref{eq:likeweights} such that the fractional change $r_i$ defined in Eq.~\eqref{eq:fractionalimprovement} is $\sim$ 2-3 for all parameters. Simultaneously, the sample mean %expectation on 
of the beyond-vacuum-GR parameter $\bsvarphi$ is within 35\% of the injection. From Eqs.~\eqref{eq:impeff} and~\eqref{eq:ess}, the importance sampling efficiency is $\eta \approx 0.08$, typical of the importance sampling class of methods for high SNR sources (see, e.g.,~\cite{Romero-Shaw:2021ual} and~\cite{Saleh:2024tgr}). The Savage-Dickey ratio, calculated from Eq.~\eqref{eq:savagedickey} is $\mathcal{B}_0^1 \approx 2.0$ in favor of the superset beyond-vacuum-GR hypothesis.\footnote{In a separate analysis with a null injection ($A=0$ in the signal), the Savage-Dickey ratio was estimated as $\approx 0.7$ in both implementations, which can be treated as the baseline for hypothesis testing.} Finally, we obtain the reweighted importance distribution by drawing $N_{\rm ess}$ samples without replacement from the set of proposed samples in proportion to their weights. The importance resamples are plotted in the top panel of Fig.~\ref{fig:5paramresamples}. We overlay our results with MCMC samples from the posterior in the true (superset) hypothesis $\mathcal{H}_1$ for comparison, which required $\sim 5 \times 10^5$ likelihood calls, making MCMC inference an order-of-magnitude more expensive than bias-corrected importance sampling in this example. Notably, the coverage is partial, showcasing a setback of the basic implementation for posterior recovery. This could result from numerical errors in the calculation or inversion of the FIM, which is especially likely for the ill-posed (highly correlated) EMRI parametrization, or if the vacuum-GR MLE estimator is incorrect, etc. (see Fig.~\ref{fig:regularizedschematic} and discussion in Sec.~\ref{sec:regularizedimplementation}).

Addressing such errors in the basic implementation is difficult, particularly without prior knowledge of the true posterior distribution, prompting a switch to the more conservative regularized implementation. Here, $A$ is chosen from a linear grid of $m = 5$ values on the prior range, and $\Sigma_{\rm reg}$ is obtained from Eqs.~\eqref{eq:regularize},~\eqref{eq:regularizationfactor}-\eqref{eq:iterativeD}, implemented iteratively until $\epsilon_k < \epsilon_0 = 10^{-12}$. Then, the same number of $\approx 5 \times 10^4$ samples $\{\bspsi',\bsvarphi'\}\sim q_{\rm reg}(\bspsi',\bsvarphi')$ are generated following section~\ref{sec:regularizedimplementation}. The importance expectations on the vacuum-GR parameters can be calculated from the sample likelihoods, such that $r_i$ (Eq.~\eqref{eq:fractionalimprovement}) is now $\sim 10$ for all parameters, much higher than $r_i$ in the basic implementation. The beyond-vacuum-GR parameter is also recovered within $7.5\%$ of the injection. In contrast, the importance sampling efficiency is lower, but is still $\eta \approx 0.05$. This is expected since the regularized proposal pdf is broader by construction, covering a larger ``empty'' region around the true posterior. The Savage-Dickey ratio in the regularized implementation is $\mathcal{B}_0^1 \approx 1.7$, 
which is consistent with the basic implementation for model selection, within sampling uncertainties. We also recover an improved coverage of the posterior pdf, as shown in the bottom panel of Fig.~\ref{fig:5paramresamples}. Thus, despite the lower importance efficiency, importance sampling under the regularized framework provides an overall improved inference of the target distribution.

\subsection{Example \rom{2}: Robustness in Higher-Dimensions}~\label{sec:highdimexample}

The extrinsic parameters are restored in the second example, as would be the case in more realistic analyses. Here, $\bspsi := (\ln M,\ln\mu, a, p_0, \Phi_0,D_L,\theta_S,\phi_S,\theta_K,\phi_K)$ and $\bsvarphi := A$ with true value $A = 3.84\times 10^{-5}$ in the signal. The perturbative effect amplitude is doubled to induce a similarly large $\sim 2\sigma$ bias on the parameters for a fair comparison (see Fig.~\ref{fig:10paramMCMC} of Appendix~\ref{app:10paramT1} for the full posterior corner plot). Again, discarding the first $99\%$ of the $\approx 1.3\times 10^6$ total MCMC samples in $\mathcal{H}_0$, and setting $m=5$, we use $N\times m \approx 6.5\times 10^4$ vacuum-GR samples in the inference. The prior range on $A$ is unchanged.

In the basic implementation, calculating the importance expectation on $\bspsi$, the fractional change $r_i$ (Eq.~\eqref{eq:fractionalimprovement}) is $\sim 2$--$5$, with the sample mean of $\bsvarphi$ within $15 \%$ of the injection. The importance efficiency is $\approx 0.05$, and the importance resamples still poorly overlap with the true posterior in the superset beyond-vacuum-GR hypothesis. The Savage-Dickey ratio is obtained following Eq.~\eqref{eq:savagedickey} as $\mathcal{B}_0^1 \approx 1.7$. In the regularized implementation, the importance samples for the same injection provide  $r_i$ for $\bspsi$ of the order $\sim 10-100$, and the beyond-vacuum-GR parameter $\bsvarphi$ is recovered within $5\%$ of the injected value. However, the importance efficiency drops to $\approx 0.01$ because of the coverage of a larger empty space in the added dimensions. The Savage-Dickey ratio in the regularized scheme is $\mathcal{B}_0^1 \approx 1.83$, again consistent with the basic implementation for model selection. The importance resamples strongly overlap with MCMC samples drawn from the posterior pdf in the true (superset) hypothesis, as shown in Fig.~\ref{fig:10paramresamples} in the Appendix. For comparison, the MCMC inference required $\sim 4 \times 10^5$ likelihood calls, making it a factor $\sim 8$ more expensive than bias-corrected importance sampling. This showcases the effectiveness of our method even in high-dimensional cases where MCMC techniques are considered the only feasible option.

%%%%%%%%%%%%%%%%%%%%%%%%%%%%%%%%%%%%%%%%%%%%%
\section{Discussion}\label{sec:discussion}
%%%%%%%%%%%%%%%%%%%%%%%%%%%%%%%%%%%%%%%%%%%%%

\subsection{Summary and outlook} 

While dynamic inference methods like MCMC are widely employed in GW data analysis, they may become computationally infeasible for the inference and comparison of alternate hypotheses in a nested framework, which includes the general class of perturbative beyond-vacuum-GR effects in binary GW sources. In this paper, we presented the bias-corrected importance sampling formalism and describe two implementations---basic and regularized---targeting the efficient inference of such effects. Empirically, we found that redoing MCMC inference of the posterior surface in the superset hypothesis required an order-of-magnitude more calls to the likelihood function than needed for bias-corrected importance sampling. Additionally, since all samples from the proposal pdf are simultaneously available in importance sampling, likelihood calls can be made parallelly. This is an inherent limitation of sequential techniques like MCMC and nested sampling. In other words, the bias-corrected importance sampling method enables the inference of $\gtrsim$ an order-of-magnitude more alternate nested hypotheses compared to methods like MCMC at the same computational expense; this is especially relevant, e.g., for high-precision probes of strong gravity sources in galactic centers like EMRIs and MBHBs, future observations~\cite{Gupta:2024gun} of which may be used to perform multiple different tests of astrophysical and modified-GR theories.

While the basic implementation underscores fundamental aspects of the bias-corrected importance sampling framework, the method heavily relies on the accuracy of the correction axis constructed following Eq.~\eqref{eq:transformationfunction}, which is susceptible to both theoretical and numerical errors, i.e., if the linear bias approximation (Eq.~\eqref{eq:cvnested}) is invalid or components thereof are incorrectly computed. In our examples, this translated to poor overlap of the importance resamples with the true posterior. The more conservative proposal defined in the regularized implementation was found to be relatively robust, providing better overall inference of the underlying posterior. Alternatively, a better-posed EMRI parametrization that promotes more stable results may resolve some of these practical challenges and will be explored in a separate work. More fundamentally, the method's effectiveness is limited by the validity of the linear-signal approximation itself, which will generally not hold for signals below a threshold SNR, typically set as $\approx 20$ in the literature (see, e.g.,~\cite{Finn:1992wt,Cutler:1994ys,Vallisneri:2007ev}). Additionally, while the dimensionality of the posterior space did not impact inference in the explored examples, the effective sample size of the importance resamples dropped considerably in the higher-dimensional example (Section~\ref{sec:highdimexample}). Thus, the inference of even higher-dimensional models, describing e.g. generic EMRI orbits, more complex beyond-vacuum-GR modifications, or combinations thereof, may require a larger starting sample size $N$.

\subsection{Future directions}

The mathematical framework developed in section~\ref{sec:methods} generically applies to any number of induced beyond-vacuum-GR effects, each with an arbitrary number of model parameters. As argued in \cite{Kejriwal:2023djc}, multiple such effects may be present in the signal simultaneously, motivating a joint inference analysis. Such combinations of beyond-vacuum-GR effects can be readily studied in follow-up work under the bias-corrected framework. Similarly, model-agnostic tests of GR (developed, e.g., in the ppE formalism~\cite{Yunes:2009ke}, or for parametrized tests of GR with observations from ground-based detectors~\cite{Li:2011cg,Saleem:2021nsb}) may be defined as the superset model and studied inexpensively in our framework. However, their parametric approach with non-physical degrees of freedom has been previously challenged~\cite{Chua:2020oxn}. Finally, we stress that our formalism is not restricted to nested model setups, and can be generically applied to infer bias-corrected targets by suitably replacing the transformation function (Eq.~\eqref{eq:transformationfunction}).

The global-fit pipeline, adopted as the primary inference pipeline for LISA data analysis~\cite{Vallisneri:2008ye,Littenberg:2023xpl,Katz:2024oqg}, aims to simultaneously infer all sources in the data stream through techniques like reversible-jump MCMC~\cite{Green:1995mxx} and blocked Gibbs sampling~\cite{f47e9344-75dc-3782-a811-9e51a952e8bf,10.5555/1571802}. In such a setup, inferring beyond-vacuum-GR effects using conventional methods may not even be feasible, since the introduction of a single beyond-vacuum-GR effect with dimensionality $d_{\bsvarphi}$ would require a new global-fit on a posterior space with $S\times d_{\bsvarphi}$ \textit{additional} dimensions, where $S$ is the number of detected sources. Alternatively, given only the vacuum-GR posteriors as the output of the global-fit analysis, our framework can readily construct proposals spanning the underlying posteriors in the beyond-vacuum-GR hypothesis, feasibly infer them, and even inform the global-fit pipeline as a consequence --- allowing robust and systematic inference of perturbative beyond-vacuum-GR effects in GW sources in the future.

\acknowledgments
We thank the organizers of the 15th International LISA Symposium, held in Dublin, Ireland, from 7-12 July 2024 where this idea was initially conceived. SK acknowledges the computing resources accessed from NUS IT Research Computing group and the support of the NUS Research Scholarship.

\appendix

\section{Example \rom{2} corner plots}\label{app:10paramT1}
\begin{figure*}
    \centering
    \includegraphics[width=0.95\linewidth]{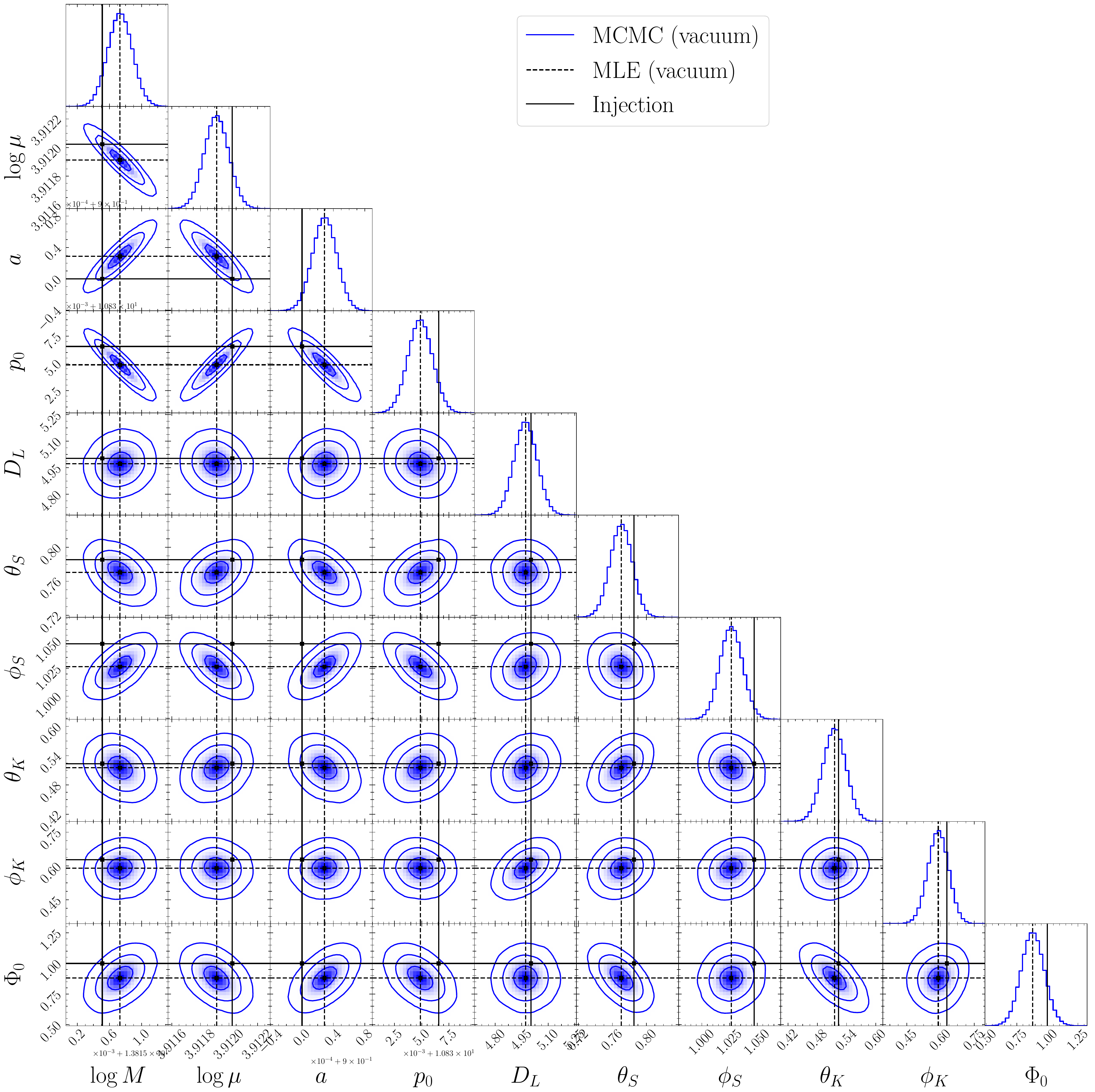}
    \caption{\justifying Distribution of posterior samples of an EMRI signal assuming the vacuum-GR (subset) hypothesis $\mathcal{H}_0$ with model parameters $\bspsi = (\ln M,\ln\mu, a, p_0, \Phi_0,D_L,\theta_S,\phi_S,\theta_K,\phi_K)$ (blue contours). The injected signal (solid line) includes an additional beyond-vacuum-GR parameter $\bsvarphi = A = 3.84\times 10^{-5}$ induced by the effect of an accretion disk around the MBH. The recovered MLE $\bspsi_{\rm MLE}|\mathcal{H}_0$ (dashed line) is $\sim 2\sigma$ biased to the injection for all parameters except $(D_L, \theta_K,\phi_K)$.}
    \label{fig:10paramMCMC}
\end{figure*}

\begin{figure*}
    \centering
    \includegraphics[width=0.95\linewidth]{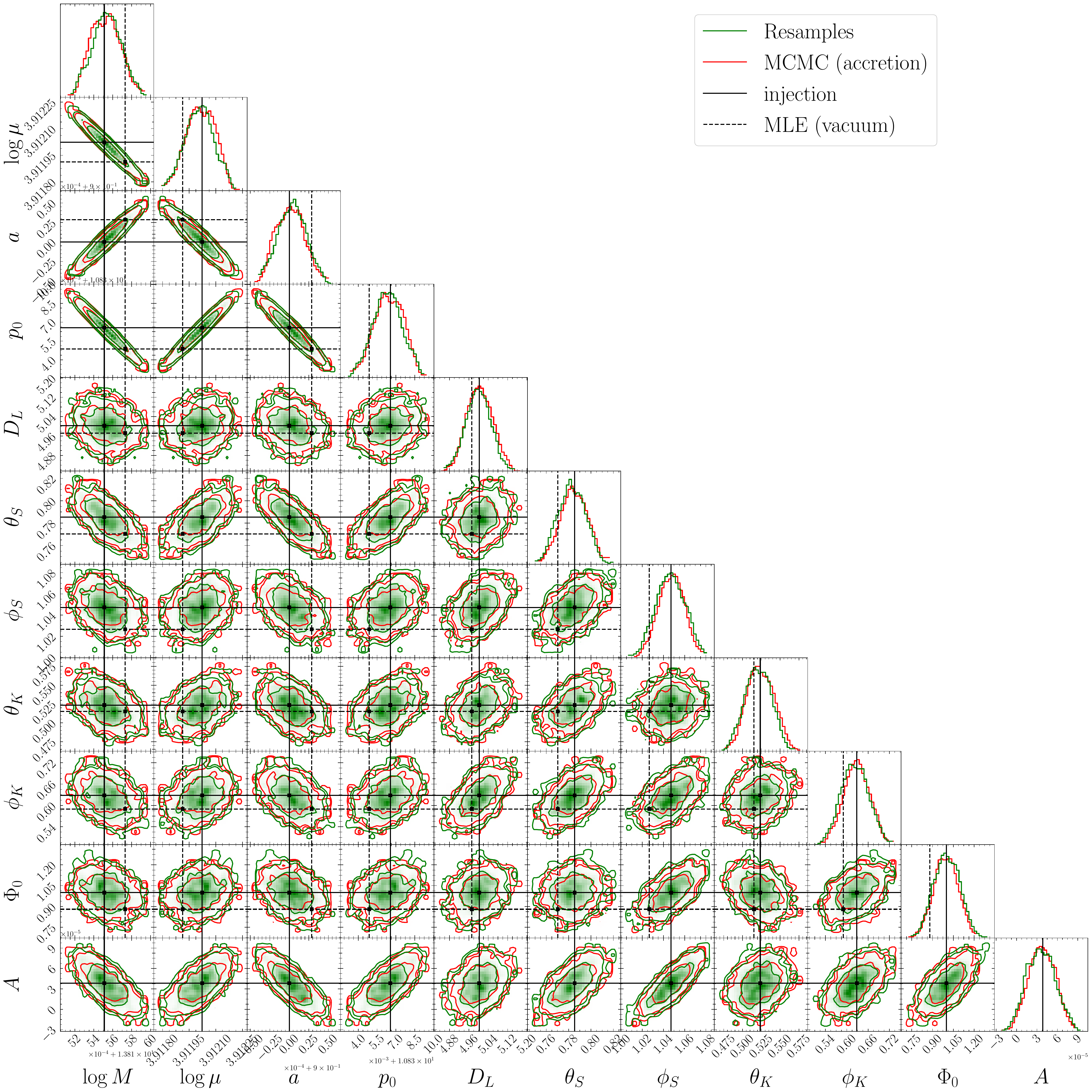}
    \caption{\justifying Distribution of bias-corrected importance \textit{re}samples (green contours) in the regularized implementation of Example \rom{2}, overlaid with posterior samples of the injected EMRI signal in the true beyond-vacuum-GR (superset) hypothesis $\mathcal{H}_1$ (red contours). %Despite the low efficiency, the importance resamples cover the posterior pdf reasonably well.
    }
    \label{fig:10paramresamples}
\end{figure*}

In Fig.~\ref{fig:10paramMCMC}, we present the full 10-dimensional posterior distribution obtained via the MCMC inference of the injected signal in Example \rom{2} assuming the vacuum-GR hypothesis. We show the corresponding importance resamples for the regularized bias-corrected importance sampling implementation in Fig.~\ref{fig:10paramresamples}, overlaid with MCMC samples from the posterior pdf in the true (superset) hypothesis.

\bibliography{references}

%merlin.mbs apsrev4-1.bst 2010-07-25 4.21a (PWD, AO, DPC) hacked
%Control: key (0)
%Control: author (8) initials jnrlst
%Control: editor formatted (1) identically to author
%Control: production of article title (-1) disabled
%Control: page (0) single
%Control: year (1) truncated
%Control: production of eprint (0) enabled
\begin{thebibliography}{73}%
\makeatletter
\providecommand \@ifxundefined [1]{%
 \@ifx{#1\undefined}
}%
\providecommand \@ifnum [1]{%
 \ifnum #1\expandafter \@firstoftwo
 \else \expandafter \@secondoftwo
 \fi
}%
\providecommand \@ifx [1]{%
 \ifx #1\expandafter \@firstoftwo
 \else \expandafter \@secondoftwo
 \fi
}%
\providecommand \natexlab [1]{#1}%
\providecommand \enquote  [1]{``#1''}%
\providecommand \bibnamefont  [1]{#1}%
\providecommand \bibfnamefont [1]{#1}%
\providecommand \citenamefont [1]{#1}%
\providecommand \href@noop [0]{\@secondoftwo}%
\providecommand \href [0]{\begingroup \@sanitize@url \@href}%
\providecommand \@href[1]{\@@startlink{#1}\@@href}%
\providecommand \@@href[1]{\endgroup#1\@@endlink}%
\providecommand \@sanitize@url [0]{\catcode `\\12\catcode `\$12\catcode `\&12\catcode `\#12\catcode `\^12\catcode `\_12\catcode `\%12\relax}%
\providecommand \@@startlink[1]{}%
\providecommand \@@endlink[0]{}%
\providecommand \url  [0]{\begingroup\@sanitize@url \@url }%
\providecommand \@url [1]{\endgroup\@href {#1}{\urlprefix }}%
\providecommand \urlprefix  [0]{URL }%
\providecommand \Eprint [0]{\href }%
\providecommand \doibase [0]{http://dx.doi.org/}%
\providecommand \selectlanguage [0]{\@gobble}%
\providecommand \bibinfo  [0]{\@secondoftwo}%
\providecommand \bibfield  [0]{\@secondoftwo}%
\providecommand \translation [1]{[#1]}%
\providecommand \BibitemOpen [0]{}%
\providecommand \bibitemStop [0]{}%
\providecommand \bibitemNoStop [0]{.\EOS\space}%
\providecommand \EOS [0]{\spacefactor3000\relax}%
\providecommand \BibitemShut  [1]{\csname bibitem#1\endcsname}%
\let\auto@bib@innerbib\@empty
%</preamble>
\bibitem [{\citenamefont {Amaro-Seoane}\ \emph {et~al.}(2017)\citenamefont {Amaro-Seoane} \emph {et~al.}}]{LISA:2017pwj}%
  \BibitemOpen
  \bibfield  {author} {\bibinfo {author} {\bibfnamefont {P.}~\bibnamefont {Amaro-Seoane}} \emph {et~al.} (\bibinfo {collaboration} {LISA}),\ }\href@noop {} {\  (\bibinfo {year} {2017})},\ \Eprint {http://arxiv.org/abs/1702.00786} {arXiv:1702.00786 [astro-ph.IM]} \BibitemShut {NoStop}%
\bibitem [{\citenamefont {Seoane}\ \emph {et~al.}(2023)\citenamefont {Seoane} \emph {et~al.}}]{LISA:2022yao}%
  \BibitemOpen
  \bibfield  {author} {\bibinfo {author} {\bibfnamefont {P.~A.}\ \bibnamefont {Seoane}} \emph {et~al.} (\bibinfo {collaboration} {LISA}),\ }\href {\doibase 10.1007/s41114-022-00041-y} {\bibfield  {journal} {\bibinfo  {journal} {Living Rev. Rel.}\ }\textbf {\bibinfo {volume} {26}},\ \bibinfo {pages} {2} (\bibinfo {year} {2023})},\ \Eprint {http://arxiv.org/abs/2203.06016} {arXiv:2203.06016 [gr-qc]} \BibitemShut {NoStop}%
\bibitem [{\citenamefont {Colpi}\ \emph {et~al.}(2024)\citenamefont {Colpi} \emph {et~al.}}]{Colpi:2024xhw}%
  \BibitemOpen
  \bibfield  {author} {\bibinfo {author} {\bibfnamefont {M.}~\bibnamefont {Colpi}} \emph {et~al.},\ }\href@noop {} {\  (\bibinfo {year} {2024})},\ \Eprint {http://arxiv.org/abs/2402.07571} {arXiv:2402.07571 [astro-ph.CO]} \BibitemShut {NoStop}%
\bibitem [{\citenamefont {Barack}\ and\ \citenamefont {Cutler}(2004)}]{Barack:2003fp}%
  \BibitemOpen
  \bibfield  {author} {\bibinfo {author} {\bibfnamefont {L.}~\bibnamefont {Barack}}\ and\ \bibinfo {author} {\bibfnamefont {C.}~\bibnamefont {Cutler}},\ }\href {\doibase 10.1103/PhysRevD.69.082005} {\bibfield  {journal} {\bibinfo  {journal} {Phys. Rev. D}\ }\textbf {\bibinfo {volume} {69}},\ \bibinfo {pages} {082005} (\bibinfo {year} {2004})},\ \Eprint {http://arxiv.org/abs/gr-qc/0310125} {arXiv:gr-qc/0310125} \BibitemShut {NoStop}%
\bibitem [{\citenamefont {Babak}\ \emph {et~al.}(2017)\citenamefont {Babak}, \citenamefont {Gair}, \citenamefont {Sesana}, \citenamefont {Barausse}, \citenamefont {Sopuerta}, \citenamefont {Berry}, \citenamefont {Berti}, \citenamefont {Amaro-Seoane}, \citenamefont {Petiteau},\ and\ \citenamefont {Klein}}]{Babak:2017tow}%
  \BibitemOpen
  \bibfield  {author} {\bibinfo {author} {\bibfnamefont {S.}~\bibnamefont {Babak}}, \bibinfo {author} {\bibfnamefont {J.}~\bibnamefont {Gair}}, \bibinfo {author} {\bibfnamefont {A.}~\bibnamefont {Sesana}}, \bibinfo {author} {\bibfnamefont {E.}~\bibnamefont {Barausse}}, \bibinfo {author} {\bibfnamefont {C.~F.}\ \bibnamefont {Sopuerta}}, \bibinfo {author} {\bibfnamefont {C.~P.~L.}\ \bibnamefont {Berry}}, \bibinfo {author} {\bibfnamefont {E.}~\bibnamefont {Berti}}, \bibinfo {author} {\bibfnamefont {P.}~\bibnamefont {Amaro-Seoane}}, \bibinfo {author} {\bibfnamefont {A.}~\bibnamefont {Petiteau}}, \ and\ \bibinfo {author} {\bibfnamefont {A.}~\bibnamefont {Klein}},\ }\href {\doibase 10.1103/PhysRevD.95.103012} {\bibfield  {journal} {\bibinfo  {journal} {Phys. Rev. D}\ }\textbf {\bibinfo {volume} {95}},\ \bibinfo {pages} {103012} (\bibinfo {year} {2017})},\ \Eprint {http://arxiv.org/abs/1703.09722} {arXiv:1703.09722 [gr-qc]} \BibitemShut {NoStop}%
\bibitem [{\citenamefont {Berry}\ \emph {et~al.}(2019)\citenamefont {Berry}, \citenamefont {Hughes}, \citenamefont {Sopuerta}, \citenamefont {Chua}, \citenamefont {Heffernan}, \citenamefont {Holley-Bockelmann}, \citenamefont {Mihaylov}, \citenamefont {Miller},\ and\ \citenamefont {Sesana}}]{Berry:2019wgg}%
  \BibitemOpen
  \bibfield  {author} {\bibinfo {author} {\bibfnamefont {C.~P.~L.}\ \bibnamefont {Berry}}, \bibinfo {author} {\bibfnamefont {S.~A.}\ \bibnamefont {Hughes}}, \bibinfo {author} {\bibfnamefont {C.~F.}\ \bibnamefont {Sopuerta}}, \bibinfo {author} {\bibfnamefont {A.~J.~K.}\ \bibnamefont {Chua}}, \bibinfo {author} {\bibfnamefont {A.}~\bibnamefont {Heffernan}}, \bibinfo {author} {\bibfnamefont {K.}~\bibnamefont {Holley-Bockelmann}}, \bibinfo {author} {\bibfnamefont {D.~P.}\ \bibnamefont {Mihaylov}}, \bibinfo {author} {\bibfnamefont {M.~C.}\ \bibnamefont {Miller}}, \ and\ \bibinfo {author} {\bibfnamefont {A.}~\bibnamefont {Sesana}},\ }\href@noop {} {\  (\bibinfo {year} {2019})},\ \Eprint {http://arxiv.org/abs/1903.03686} {arXiv:1903.03686 [astro-ph.HE]} \BibitemShut {NoStop}%
\bibitem [{\citenamefont {Yunes}\ and\ \citenamefont {Pretorius}(2009)}]{Yunes:2009ke}%
  \BibitemOpen
  \bibfield  {author} {\bibinfo {author} {\bibfnamefont {N.}~\bibnamefont {Yunes}}\ and\ \bibinfo {author} {\bibfnamefont {F.}~\bibnamefont {Pretorius}},\ }\href {\doibase 10.1103/PhysRevD.80.122003} {\bibfield  {journal} {\bibinfo  {journal} {Phys. Rev. D}\ }\textbf {\bibinfo {volume} {80}},\ \bibinfo {pages} {122003} (\bibinfo {year} {2009})},\ \Eprint {http://arxiv.org/abs/0909.3328} {arXiv:0909.3328 [gr-qc]} \BibitemShut {NoStop}%
\bibitem [{\citenamefont {Barausse}\ \emph {et~al.}(2016)\citenamefont {Barausse}, \citenamefont {Yunes},\ and\ \citenamefont {Chamberlain}}]{Barausse:2016eii}%
  \BibitemOpen
  \bibfield  {author} {\bibinfo {author} {\bibfnamefont {E.}~\bibnamefont {Barausse}}, \bibinfo {author} {\bibfnamefont {N.}~\bibnamefont {Yunes}}, \ and\ \bibinfo {author} {\bibfnamefont {K.}~\bibnamefont {Chamberlain}},\ }\href {\doibase 10.1103/PhysRevLett.116.241104} {\bibfield  {journal} {\bibinfo  {journal} {Phys. Rev. Lett.}\ }\textbf {\bibinfo {volume} {116}},\ \bibinfo {pages} {241104} (\bibinfo {year} {2016})},\ \Eprint {http://arxiv.org/abs/1603.04075} {arXiv:1603.04075 [gr-qc]} \BibitemShut {NoStop}%
\bibitem [{\citenamefont {Arun}\ \emph {et~al.}(2022)\citenamefont {Arun} \emph {et~al.}}]{LISA:2022kgy}%
  \BibitemOpen
  \bibfield  {author} {\bibinfo {author} {\bibfnamefont {K.~G.}\ \bibnamefont {Arun}} \emph {et~al.} (\bibinfo {collaboration} {LISA}),\ }\href {\doibase 10.1007/s41114-022-00036-9} {\bibfield  {journal} {\bibinfo  {journal} {Living Rev. Rel.}\ }\textbf {\bibinfo {volume} {25}},\ \bibinfo {pages} {4} (\bibinfo {year} {2022})},\ \Eprint {http://arxiv.org/abs/2205.01597} {arXiv:2205.01597 [gr-qc]} \BibitemShut {NoStop}%
\bibitem [{\citenamefont {Speri}\ \emph {et~al.}(2024)\citenamefont {Speri}, \citenamefont {Barsanti}, \citenamefont {Maselli}, \citenamefont {Sotiriou}, \citenamefont {Warburton}, \citenamefont {van~de Meent}, \citenamefont {Chua}, \citenamefont {Burke},\ and\ \citenamefont {Gair}}]{Speri:2024qak}%
  \BibitemOpen
  \bibfield  {author} {\bibinfo {author} {\bibfnamefont {L.}~\bibnamefont {Speri}}, \bibinfo {author} {\bibfnamefont {S.}~\bibnamefont {Barsanti}}, \bibinfo {author} {\bibfnamefont {A.}~\bibnamefont {Maselli}}, \bibinfo {author} {\bibfnamefont {T.~P.}\ \bibnamefont {Sotiriou}}, \bibinfo {author} {\bibfnamefont {N.}~\bibnamefont {Warburton}}, \bibinfo {author} {\bibfnamefont {M.}~\bibnamefont {van~de Meent}}, \bibinfo {author} {\bibfnamefont {A.~J.~K.}\ \bibnamefont {Chua}}, \bibinfo {author} {\bibfnamefont {O.}~\bibnamefont {Burke}}, \ and\ \bibinfo {author} {\bibfnamefont {J.}~\bibnamefont {Gair}},\ }\href@noop {} {\  (\bibinfo {year} {2024})},\ \Eprint {http://arxiv.org/abs/2406.07607} {arXiv:2406.07607 [gr-qc]} \BibitemShut {NoStop}%
\bibitem [{\citenamefont {Barausse}\ \emph {et~al.}(2014)\citenamefont {Barausse}, \citenamefont {Cardoso},\ and\ \citenamefont {Pani}}]{Barausse:2014tra}%
  \BibitemOpen
  \bibfield  {author} {\bibinfo {author} {\bibfnamefont {E.}~\bibnamefont {Barausse}}, \bibinfo {author} {\bibfnamefont {V.}~\bibnamefont {Cardoso}}, \ and\ \bibinfo {author} {\bibfnamefont {P.}~\bibnamefont {Pani}},\ }\href {\doibase 10.1103/PhysRevD.89.104059} {\bibfield  {journal} {\bibinfo  {journal} {Phys. Rev. D}\ }\textbf {\bibinfo {volume} {89}},\ \bibinfo {pages} {104059} (\bibinfo {year} {2014})},\ \Eprint {http://arxiv.org/abs/1404.7149} {arXiv:1404.7149 [gr-qc]} \BibitemShut {NoStop}%
\bibitem [{\citenamefont {Speri}\ \emph {et~al.}(2023)\citenamefont {Speri}, \citenamefont {Antonelli}, \citenamefont {Sberna}, \citenamefont {Babak}, \citenamefont {Barausse}, \citenamefont {Gair},\ and\ \citenamefont {Katz}}]{Speri:2022upm}%
  \BibitemOpen
  \bibfield  {author} {\bibinfo {author} {\bibfnamefont {L.}~\bibnamefont {Speri}}, \bibinfo {author} {\bibfnamefont {A.}~\bibnamefont {Antonelli}}, \bibinfo {author} {\bibfnamefont {L.}~\bibnamefont {Sberna}}, \bibinfo {author} {\bibfnamefont {S.}~\bibnamefont {Babak}}, \bibinfo {author} {\bibfnamefont {E.}~\bibnamefont {Barausse}}, \bibinfo {author} {\bibfnamefont {J.~R.}\ \bibnamefont {Gair}}, \ and\ \bibinfo {author} {\bibfnamefont {M.~L.}\ \bibnamefont {Katz}},\ }\href {\doibase 10.1103/PhysRevX.13.021035} {\bibfield  {journal} {\bibinfo  {journal} {Phys. Rev. X}\ }\textbf {\bibinfo {volume} {13}},\ \bibinfo {pages} {021035} (\bibinfo {year} {2023})},\ \Eprint {http://arxiv.org/abs/2207.10086} {arXiv:2207.10086 [gr-qc]} \BibitemShut {NoStop}%
\bibitem [{\citenamefont {Kocsis}\ \emph {et~al.}(2011)\citenamefont {Kocsis}, \citenamefont {Yunes},\ and\ \citenamefont {Loeb}}]{Kocsis:2011dr}%
  \BibitemOpen
  \bibfield  {author} {\bibinfo {author} {\bibfnamefont {B.}~\bibnamefont {Kocsis}}, \bibinfo {author} {\bibfnamefont {N.}~\bibnamefont {Yunes}}, \ and\ \bibinfo {author} {\bibfnamefont {A.}~\bibnamefont {Loeb}},\ }\href {\doibase 10.1103/PhysRevD.86.049907} {\bibfield  {journal} {\bibinfo  {journal} {Phys. Rev. D}\ }\textbf {\bibinfo {volume} {84}},\ \bibinfo {pages} {024032} (\bibinfo {year} {2011})},\ \Eprint {http://arxiv.org/abs/1104.2322} {arXiv:1104.2322 [astro-ph.GA]} \BibitemShut {NoStop}%
\bibitem [{\citenamefont {Garg}\ \emph {et~al.}(2024{\natexlab{a}})\citenamefont {Garg}, \citenamefont {Sberna}, \citenamefont {Speri}, \citenamefont {Duque},\ and\ \citenamefont {Gair}}]{Garg:2024qxq}%
  \BibitemOpen
  \bibfield  {author} {\bibinfo {author} {\bibfnamefont {M.}~\bibnamefont {Garg}}, \bibinfo {author} {\bibfnamefont {L.}~\bibnamefont {Sberna}}, \bibinfo {author} {\bibfnamefont {L.}~\bibnamefont {Speri}}, \bibinfo {author} {\bibfnamefont {F.}~\bibnamefont {Duque}}, \ and\ \bibinfo {author} {\bibfnamefont {J.}~\bibnamefont {Gair}},\ }\href@noop {} {\  (\bibinfo {year} {2024}{\natexlab{a}})},\ \Eprint {http://arxiv.org/abs/2410.02910} {arXiv:2410.02910 [astro-ph.GA]} \BibitemShut {NoStop}%
\bibitem [{\citenamefont {Duque}\ \emph {et~al.}(2024)\citenamefont {Duque}, \citenamefont {Kejriwal}, \citenamefont {Sberna}, \citenamefont {Speri},\ and\ \citenamefont {Gair}}]{Duque:2024mfw}%
  \BibitemOpen
  \bibfield  {author} {\bibinfo {author} {\bibfnamefont {F.}~\bibnamefont {Duque}}, \bibinfo {author} {\bibfnamefont {S.}~\bibnamefont {Kejriwal}}, \bibinfo {author} {\bibfnamefont {L.}~\bibnamefont {Sberna}}, \bibinfo {author} {\bibfnamefont {L.}~\bibnamefont {Speri}}, \ and\ \bibinfo {author} {\bibfnamefont {J.}~\bibnamefont {Gair}},\ }\href@noop {} {\  (\bibinfo {year} {2024})},\ \Eprint {http://arxiv.org/abs/2411.03436} {arXiv:2411.03436 [gr-qc]} \BibitemShut {NoStop}%
\bibitem [{\citenamefont {Favata}(2014)}]{Favata:2013rwa}%
  \BibitemOpen
  \bibfield  {author} {\bibinfo {author} {\bibfnamefont {M.}~\bibnamefont {Favata}},\ }\href {\doibase 10.1103/PhysRevLett.112.101101} {\bibfield  {journal} {\bibinfo  {journal} {Phys. Rev. Lett.}\ }\textbf {\bibinfo {volume} {112}},\ \bibinfo {pages} {101101} (\bibinfo {year} {2014})},\ \Eprint {http://arxiv.org/abs/1310.8288} {arXiv:1310.8288 [gr-qc]} \BibitemShut {NoStop}%
\bibitem [{\citenamefont {Romero-Shaw}\ \emph {et~al.}(2020)\citenamefont {Romero-Shaw}, \citenamefont {Lasky}, \citenamefont {Thrane},\ and\ \citenamefont {Bustillo}}]{Romero-Shaw:2020thy}%
  \BibitemOpen
  \bibfield  {author} {\bibinfo {author} {\bibfnamefont {I.~M.}\ \bibnamefont {Romero-Shaw}}, \bibinfo {author} {\bibfnamefont {P.~D.}\ \bibnamefont {Lasky}}, \bibinfo {author} {\bibfnamefont {E.}~\bibnamefont {Thrane}}, \ and\ \bibinfo {author} {\bibfnamefont {J.~C.}\ \bibnamefont {Bustillo}},\ }\href {\doibase 10.3847/2041-8213/abbe26} {\bibfield  {journal} {\bibinfo  {journal} {Astrophys. J. Lett.}\ }\textbf {\bibinfo {volume} {903}},\ \bibinfo {pages} {L5} (\bibinfo {year} {2020})},\ \Eprint {http://arxiv.org/abs/2009.04771} {arXiv:2009.04771 [astro-ph.HE]} \BibitemShut {NoStop}%
\bibitem [{\citenamefont {Romero-Shaw}\ \emph {et~al.}(2021)\citenamefont {Romero-Shaw}, \citenamefont {Lasky},\ and\ \citenamefont {Thrane}}]{Romero-Shaw:2021ual}%
  \BibitemOpen
  \bibfield  {author} {\bibinfo {author} {\bibfnamefont {I.~M.}\ \bibnamefont {Romero-Shaw}}, \bibinfo {author} {\bibfnamefont {P.~D.}\ \bibnamefont {Lasky}}, \ and\ \bibinfo {author} {\bibfnamefont {E.}~\bibnamefont {Thrane}},\ }\href {\doibase 10.3847/2041-8213/ac3138} {\bibfield  {journal} {\bibinfo  {journal} {Astrophys. J. Lett.}\ }\textbf {\bibinfo {volume} {921}},\ \bibinfo {pages} {L31} (\bibinfo {year} {2021})},\ \Eprint {http://arxiv.org/abs/2108.01284} {arXiv:2108.01284 [astro-ph.HE]} \BibitemShut {NoStop}%
\bibitem [{\citenamefont {Piovano}\ \emph {et~al.}(2020)\citenamefont {Piovano}, \citenamefont {Maselli},\ and\ \citenamefont {Pani}}]{Piovano:2020zin}%
  \BibitemOpen
  \bibfield  {author} {\bibinfo {author} {\bibfnamefont {G.~A.}\ \bibnamefont {Piovano}}, \bibinfo {author} {\bibfnamefont {A.}~\bibnamefont {Maselli}}, \ and\ \bibinfo {author} {\bibfnamefont {P.}~\bibnamefont {Pani}},\ }\href {\doibase 10.1103/PhysRevD.102.024041} {\bibfield  {journal} {\bibinfo  {journal} {Phys. Rev. D}\ }\textbf {\bibinfo {volume} {102}},\ \bibinfo {pages} {024041} (\bibinfo {year} {2020})},\ \Eprint {http://arxiv.org/abs/2004.02654} {arXiv:2004.02654 [gr-qc]} \BibitemShut {NoStop}%
\bibitem [{\citenamefont {Mathews}\ \emph {et~al.}(2022)\citenamefont {Mathews}, \citenamefont {Pound},\ and\ \citenamefont {Wardell}}]{Mathews:2021rod}%
  \BibitemOpen
  \bibfield  {author} {\bibinfo {author} {\bibfnamefont {J.}~\bibnamefont {Mathews}}, \bibinfo {author} {\bibfnamefont {A.}~\bibnamefont {Pound}}, \ and\ \bibinfo {author} {\bibfnamefont {B.}~\bibnamefont {Wardell}},\ }\href {\doibase 10.1103/PhysRevD.105.084031} {\bibfield  {journal} {\bibinfo  {journal} {Phys. Rev. D}\ }\textbf {\bibinfo {volume} {105}},\ \bibinfo {pages} {084031} (\bibinfo {year} {2022})},\ \Eprint {http://arxiv.org/abs/2112.13069} {arXiv:2112.13069 [gr-qc]} \BibitemShut {NoStop}%
\bibitem [{\citenamefont {Skoup\'y}\ and\ \citenamefont {Witzany}(2024)}]{Skoupy:2024uan}%
  \BibitemOpen
  \bibfield  {author} {\bibinfo {author} {\bibfnamefont {V.}~\bibnamefont {Skoup\'y}}\ and\ \bibinfo {author} {\bibfnamefont {V.}~\bibnamefont {Witzany}},\ }\href@noop {} {\  (\bibinfo {year} {2024})},\ \Eprint {http://arxiv.org/abs/2411.16855} {arXiv:2411.16855 [gr-qc]} \BibitemShut {NoStop}%
\bibitem [{\citenamefont {Lyu}\ \emph {et~al.}(2024)\citenamefont {Lyu}, \citenamefont {Pan}, \citenamefont {Mao}, \citenamefont {Jiang},\ and\ \citenamefont {Yang}}]{Lyu:2024gnk}%
  \BibitemOpen
  \bibfield  {author} {\bibinfo {author} {\bibfnamefont {Z.}~\bibnamefont {Lyu}}, \bibinfo {author} {\bibfnamefont {Z.}~\bibnamefont {Pan}}, \bibinfo {author} {\bibfnamefont {J.}~\bibnamefont {Mao}}, \bibinfo {author} {\bibfnamefont {N.}~\bibnamefont {Jiang}}, \ and\ \bibinfo {author} {\bibfnamefont {H.}~\bibnamefont {Yang}},\ }\href@noop {} {\  (\bibinfo {year} {2024})},\ \Eprint {http://arxiv.org/abs/2501.03252} {arXiv:2501.03252 [astro-ph.HE]} \BibitemShut {NoStop}%
\bibitem [{\citenamefont {Hinderer}\ and\ \citenamefont {Flanagan}(2008)}]{Hinderer:2008dm}%
  \BibitemOpen
  \bibfield  {author} {\bibinfo {author} {\bibfnamefont {T.}~\bibnamefont {Hinderer}}\ and\ \bibinfo {author} {\bibfnamefont {E.~E.}\ \bibnamefont {Flanagan}},\ }\href {\doibase 10.1103/PhysRevD.78.064028} {\bibfield  {journal} {\bibinfo  {journal} {Phys. Rev. D}\ }\textbf {\bibinfo {volume} {78}},\ \bibinfo {pages} {064028} (\bibinfo {year} {2008})},\ \Eprint {http://arxiv.org/abs/0805.3337} {arXiv:0805.3337 [gr-qc]} \BibitemShut {NoStop}%
\bibitem [{\citenamefont {Barack}\ and\ \citenamefont {Pound}(2019)}]{Barack:2018yvs}%
  \BibitemOpen
  \bibfield  {author} {\bibinfo {author} {\bibfnamefont {L.}~\bibnamefont {Barack}}\ and\ \bibinfo {author} {\bibfnamefont {A.}~\bibnamefont {Pound}},\ }\href {\doibase 10.1088/1361-6633/aae552} {\bibfield  {journal} {\bibinfo  {journal} {Rept. Prog. Phys.}\ }\textbf {\bibinfo {volume} {82}},\ \bibinfo {pages} {016904} (\bibinfo {year} {2019})},\ \Eprint {http://arxiv.org/abs/1805.10385} {arXiv:1805.10385 [gr-qc]} \BibitemShut {NoStop}%
\bibitem [{\citenamefont {Pound}\ and\ \citenamefont {Wardell}(2021)}]{Pound:2021qin}%
  \BibitemOpen
  \bibfield  {author} {\bibinfo {author} {\bibfnamefont {A.}~\bibnamefont {Pound}}\ and\ \bibinfo {author} {\bibfnamefont {B.}~\bibnamefont {Wardell}},\ }\href {\doibase {10.1007/978-981-15-4702-7\_38-1}} {\  (\bibinfo {year} {2021}),\ {10.1007/978-981-15-4702-7\_38-1}},\ \Eprint {http://arxiv.org/abs/2101.04592} {arXiv:2101.04592 [gr-qc]} \BibitemShut {NoStop}%
\bibitem [{\citenamefont {Fujita}\ and\ \citenamefont {Shibata}(2020)}]{Fujita:2020zxe}%
  \BibitemOpen
  \bibfield  {author} {\bibinfo {author} {\bibfnamefont {R.}~\bibnamefont {Fujita}}\ and\ \bibinfo {author} {\bibfnamefont {M.}~\bibnamefont {Shibata}},\ }\href {\doibase 10.1103/PhysRevD.102.064005} {\bibfield  {journal} {\bibinfo  {journal} {Phys. Rev. D}\ }\textbf {\bibinfo {volume} {102}},\ \bibinfo {pages} {064005} (\bibinfo {year} {2020})},\ \Eprint {http://arxiv.org/abs/2008.13554} {arXiv:2008.13554 [gr-qc]} \BibitemShut {NoStop}%
\bibitem [{\citenamefont {Isoyama}\ \emph {et~al.}(2022)\citenamefont {Isoyama}, \citenamefont {Fujita}, \citenamefont {Chua}, \citenamefont {Nakano}, \citenamefont {Pound},\ and\ \citenamefont {Sago}}]{Isoyama:2021jjd}%
  \BibitemOpen
  \bibfield  {author} {\bibinfo {author} {\bibfnamefont {S.}~\bibnamefont {Isoyama}}, \bibinfo {author} {\bibfnamefont {R.}~\bibnamefont {Fujita}}, \bibinfo {author} {\bibfnamefont {A.~J.~K.}\ \bibnamefont {Chua}}, \bibinfo {author} {\bibfnamefont {H.}~\bibnamefont {Nakano}}, \bibinfo {author} {\bibfnamefont {A.}~\bibnamefont {Pound}}, \ and\ \bibinfo {author} {\bibfnamefont {N.}~\bibnamefont {Sago}},\ }\href {\doibase 10.1103/PhysRevLett.128.231101} {\bibfield  {journal} {\bibinfo  {journal} {Phys. Rev. Lett.}\ }\textbf {\bibinfo {volume} {128}},\ \bibinfo {pages} {231101} (\bibinfo {year} {2022})},\ \Eprint {http://arxiv.org/abs/2111.05288} {arXiv:2111.05288 [gr-qc]} \BibitemShut {NoStop}%
\bibitem [{\citenamefont {Hughes}\ \emph {et~al.}(2021)\citenamefont {Hughes}, \citenamefont {Warburton}, \citenamefont {Khanna}, \citenamefont {Chua},\ and\ \citenamefont {Katz}}]{Hughes:2021exa}%
  \BibitemOpen
  \bibfield  {author} {\bibinfo {author} {\bibfnamefont {S.~A.}\ \bibnamefont {Hughes}}, \bibinfo {author} {\bibfnamefont {N.}~\bibnamefont {Warburton}}, \bibinfo {author} {\bibfnamefont {G.}~\bibnamefont {Khanna}}, \bibinfo {author} {\bibfnamefont {A.~J.~K.}\ \bibnamefont {Chua}}, \ and\ \bibinfo {author} {\bibfnamefont {M.~L.}\ \bibnamefont {Katz}},\ }\href {\doibase 10.1103/PhysRevD.103.104014} {\bibfield  {journal} {\bibinfo  {journal} {Phys. Rev. D}\ }\textbf {\bibinfo {volume} {103}},\ \bibinfo {pages} {104014} (\bibinfo {year} {2021})},\ \bibinfo {note} {[Erratum: Phys.Rev.D 107, 089901 (2023)]},\ \Eprint {http://arxiv.org/abs/2102.02713} {arXiv:2102.02713 [gr-qc]} \BibitemShut {NoStop}%
\bibitem [{\citenamefont {Kejriwal}\ \emph {et~al.}(2024)\citenamefont {Kejriwal}, \citenamefont {Speri},\ and\ \citenamefont {Chua}}]{Kejriwal:2023djc}%
  \BibitemOpen
  \bibfield  {author} {\bibinfo {author} {\bibfnamefont {S.}~\bibnamefont {Kejriwal}}, \bibinfo {author} {\bibfnamefont {L.}~\bibnamefont {Speri}}, \ and\ \bibinfo {author} {\bibfnamefont {A.~J.~K.}\ \bibnamefont {Chua}},\ }\href {\doibase 10.1103/PhysRevD.110.084060} {\bibfield  {journal} {\bibinfo  {journal} {Phys. Rev. D}\ }\textbf {\bibinfo {volume} {110}},\ \bibinfo {pages} {084060} (\bibinfo {year} {2024})},\ \Eprint {http://arxiv.org/abs/2312.13028} {arXiv:2312.13028 [gr-qc]} \BibitemShut {NoStop}%
\bibitem [{\citenamefont {Robert}\ and\ \citenamefont {Casella}(2004)}]{robert_monte_2004}%
  \BibitemOpen
  \bibfield  {author} {\bibinfo {author} {\bibfnamefont {C.}~\bibnamefont {Robert}}\ and\ \bibinfo {author} {\bibfnamefont {G.}~\bibnamefont {Casella}},\ }\href@noop {} {\emph {\bibinfo {title} {Monte {Carlo} statistical methods}}}\ (\bibinfo  {publisher} {Springer Verlag},\ \bibinfo {year} {2004})\BibitemShut {NoStop}%
\bibitem [{\citenamefont {Skilling}(2006)}]{Skilling:2006gxv}%
  \BibitemOpen
  \bibfield  {author} {\bibinfo {author} {\bibfnamefont {J.}~\bibnamefont {Skilling}},\ }\href {\doibase 10.1214/06-BA127} {\bibfield  {journal} {\bibinfo  {journal} {Bayesian Analysis}\ }\textbf {\bibinfo {volume} {1}},\ \bibinfo {pages} {833} (\bibinfo {year} {2006})}\BibitemShut {NoStop}%
\bibitem [{\citenamefont {{Higson}}\ \emph {et~al.}(2019)\citenamefont {{Higson}}, \citenamefont {{Handley}}, \citenamefont {{Hobson}},\ and\ \citenamefont {{Lasenby}}}]{2019S&C....29..891H}%
  \BibitemOpen
  \bibfield  {author} {\bibinfo {author} {\bibfnamefont {E.}~\bibnamefont {{Higson}}}, \bibinfo {author} {\bibfnamefont {W.}~\bibnamefont {{Handley}}}, \bibinfo {author} {\bibfnamefont {M.}~\bibnamefont {{Hobson}}}, \ and\ \bibinfo {author} {\bibfnamefont {A.}~\bibnamefont {{Lasenby}}},\ }\href {\doibase 10.1007/s11222-018-9844-0} {\bibfield  {journal} {\bibinfo  {journal} {Statistics and Computing}\ }\textbf {\bibinfo {volume} {29}},\ \bibinfo {pages} {891} (\bibinfo {year} {2019})},\ \Eprint {http://arxiv.org/abs/1704.03459} {arXiv:1704.03459 [stat.CO]} \BibitemShut {NoStop}%
\bibitem [{\citenamefont {Cutler}\ and\ \citenamefont {Vallisneri}(2007)}]{Cutler:2007mi}%
  \BibitemOpen
  \bibfield  {author} {\bibinfo {author} {\bibfnamefont {C.}~\bibnamefont {Cutler}}\ and\ \bibinfo {author} {\bibfnamefont {M.}~\bibnamefont {Vallisneri}},\ }\href {\doibase 10.1103/PhysRevD.76.104018} {\bibfield  {journal} {\bibinfo  {journal} {Phys. Rev. D}\ }\textbf {\bibinfo {volume} {76}},\ \bibinfo {pages} {104018} (\bibinfo {year} {2007})},\ \Eprint {http://arxiv.org/abs/0707.2982} {arXiv:0707.2982 [gr-qc]} \BibitemShut {NoStop}%
\bibitem [{\citenamefont {Abbott}\ \emph {et~al.}(2016)\citenamefont {Abbott} \emph {et~al.}}]{LIGOScientific:2016lio}%
  \BibitemOpen
  \bibfield  {author} {\bibinfo {author} {\bibfnamefont {B.~P.}\ \bibnamefont {Abbott}} \emph {et~al.} (\bibinfo {collaboration} {LIGO Scientific, Virgo}),\ }\href {\doibase 10.1103/PhysRevLett.116.221101} {\bibfield  {journal} {\bibinfo  {journal} {Phys. Rev. Lett.}\ }\textbf {\bibinfo {volume} {116}},\ \bibinfo {pages} {221101} (\bibinfo {year} {2016})},\ \bibinfo {note} {[Erratum: Phys.Rev.Lett. 121, 129902 (2018)]},\ \Eprint {http://arxiv.org/abs/1602.03841} {arXiv:1602.03841 [gr-qc]} \BibitemShut {NoStop}%
\bibitem [{\citenamefont {Agazie}\ \emph {et~al.}(2023{\natexlab{a}})\citenamefont {Agazie} \emph {et~al.}}]{NANOGrav:2023pdq}%
  \BibitemOpen
  \bibfield  {author} {\bibinfo {author} {\bibfnamefont {G.}~\bibnamefont {Agazie}} \emph {et~al.} (\bibinfo {collaboration} {NANOGrav}),\ }\href {\doibase 10.3847/2041-8213/ace18a} {\bibfield  {journal} {\bibinfo  {journal} {Astrophys. J. Lett.}\ }\textbf {\bibinfo {volume} {951}},\ \bibinfo {pages} {L50} (\bibinfo {year} {2023}{\natexlab{a}})},\ \Eprint {http://arxiv.org/abs/2306.16222} {arXiv:2306.16222 [astro-ph.HE]} \BibitemShut {NoStop}%
\bibitem [{\citenamefont {Gelman}\ \emph {et~al.}(2004)\citenamefont {Gelman}, \citenamefont {Carlin}, \citenamefont {Stern},\ and\ \citenamefont {Rubin}}]{gelmanbda04}%
  \BibitemOpen
  \bibfield  {author} {\bibinfo {author} {\bibfnamefont {A.}~\bibnamefont {Gelman}}, \bibinfo {author} {\bibfnamefont {J.~B.}\ \bibnamefont {Carlin}}, \bibinfo {author} {\bibfnamefont {H.~S.}\ \bibnamefont {Stern}}, \ and\ \bibinfo {author} {\bibfnamefont {D.~B.}\ \bibnamefont {Rubin}},\ }\href@noop {} {\emph {\bibinfo {title} {Bayesian Data Analysis}}},\ \bibinfo {edition} {2nd}\ ed.\ (\bibinfo  {publisher} {Chapman and Hall/CRC},\ \bibinfo {year} {2004})\BibitemShut {NoStop}%
\bibitem [{\citenamefont {Li}(2004)}]{doi:https://doi.org/10.1002/0470090456.ch24}%
  \BibitemOpen
  \bibfield  {author} {\bibinfo {author} {\bibfnamefont {K.-H.}\ \bibnamefont {Li}},\ }\enquote {\bibinfo {title} {The sampling/importance resampling algorithm},}\ in\ \href {\doibase https://doi.org/10.1002/0470090456.ch24} {\emph {\bibinfo {booktitle} {Applied Bayesian Modeling and Causal Inference from Incomplete‐Data Perspectives}}}\ (\bibinfo  {publisher} {John Wiley \& Sons, Ltd},\ \bibinfo {year} {2004})\ Chap.~\bibinfo {chapter} {24}, pp.\ \bibinfo {pages} {265--276},\ \Eprint {http://arxiv.org/abs/https://onlinelibrary.wiley.com/doi/pdf/10.1002/0470090456.ch24} {https://onlinelibrary.wiley.com/doi/pdf/10.1002/0470090456.ch24} \BibitemShut {NoStop}%
\bibitem [{\citenamefont {Jaynes}(2003)}]{jaynes03}%
  \BibitemOpen
  \bibfield  {author} {\bibinfo {author} {\bibfnamefont {E.~T.}\ \bibnamefont {Jaynes}},\ }\href@noop {} {\emph {\bibinfo {title} {Probability theory: The logic of science}}}\ (\bibinfo  {publisher} {Cambridge University Press},\ \bibinfo {address} {Cambridge},\ \bibinfo {year} {2003})\BibitemShut {NoStop}%
\bibitem [{\citenamefont {Finn}(1992)}]{Finn:1992wt}%
  \BibitemOpen
  \bibfield  {author} {\bibinfo {author} {\bibfnamefont {L.~S.}\ \bibnamefont {Finn}},\ }\href {\doibase 10.1103/PhysRevD.46.5236} {\bibfield  {journal} {\bibinfo  {journal} {Phys. Rev. D}\ }\textbf {\bibinfo {volume} {46}},\ \bibinfo {pages} {5236} (\bibinfo {year} {1992})},\ \Eprint {http://arxiv.org/abs/gr-qc/9209010} {arXiv:gr-qc/9209010} \BibitemShut {NoStop}%
\bibitem [{\citenamefont {Cutler}\ and\ \citenamefont {Flanagan}(1994)}]{Cutler:1994ys}%
  \BibitemOpen
  \bibfield  {author} {\bibinfo {author} {\bibfnamefont {C.}~\bibnamefont {Cutler}}\ and\ \bibinfo {author} {\bibfnamefont {E.~E.}\ \bibnamefont {Flanagan}},\ }\href {\doibase 10.1103/PhysRevD.49.2658} {\bibfield  {journal} {\bibinfo  {journal} {Phys. Rev. D}\ }\textbf {\bibinfo {volume} {49}},\ \bibinfo {pages} {2658} (\bibinfo {year} {1994})},\ \Eprint {http://arxiv.org/abs/gr-qc/9402014} {arXiv:gr-qc/9402014} \BibitemShut {NoStop}%
\bibitem [{\citenamefont {Abbott}\ \emph {et~al.}(2019)\citenamefont {Abbott} \emph {et~al.}}]{LIGOScientific:2019fpa}%
  \BibitemOpen
  \bibfield  {author} {\bibinfo {author} {\bibfnamefont {B.~P.}\ \bibnamefont {Abbott}} \emph {et~al.} (\bibinfo {collaboration} {LIGO Scientific, Virgo}),\ }\href {\doibase 10.1103/PhysRevD.100.104036} {\bibfield  {journal} {\bibinfo  {journal} {Phys. Rev. D}\ }\textbf {\bibinfo {volume} {100}},\ \bibinfo {pages} {104036} (\bibinfo {year} {2019})},\ \Eprint {http://arxiv.org/abs/1903.04467} {arXiv:1903.04467 [gr-qc]} \BibitemShut {NoStop}%
\bibitem [{\citenamefont {Abbott}\ \emph {et~al.}(2021{\natexlab{a}})\citenamefont {Abbott} \emph {et~al.}}]{LIGOScientific:2020tif}%
  \BibitemOpen
  \bibfield  {author} {\bibinfo {author} {\bibfnamefont {R.}~\bibnamefont {Abbott}} \emph {et~al.} (\bibinfo {collaboration} {LIGO Scientific, Virgo}),\ }\href {\doibase 10.1103/PhysRevD.103.122002} {\bibfield  {journal} {\bibinfo  {journal} {Phys. Rev. D}\ }\textbf {\bibinfo {volume} {103}},\ \bibinfo {pages} {122002} (\bibinfo {year} {2021}{\natexlab{a}})},\ \Eprint {http://arxiv.org/abs/2010.14529} {arXiv:2010.14529 [gr-qc]} \BibitemShut {NoStop}%
\bibitem [{\citenamefont {Abbott}\ \emph {et~al.}(2021{\natexlab{b}})\citenamefont {Abbott} \emph {et~al.}}]{LIGOScientific:2021sio}%
  \BibitemOpen
  \bibfield  {author} {\bibinfo {author} {\bibfnamefont {R.}~\bibnamefont {Abbott}} \emph {et~al.} (\bibinfo {collaboration} {LIGO Scientific, VIRGO, KAGRA}),\ }\href@noop {} {\  (\bibinfo {year} {2021}{\natexlab{b}})},\ \Eprint {http://arxiv.org/abs/2112.06861} {arXiv:2112.06861 [gr-qc]} \BibitemShut {NoStop}%
\bibitem [{\citenamefont {Agazie}\ \emph {et~al.}(2023{\natexlab{b}})\citenamefont {Agazie} \emph {et~al.}}]{NANOGrav:2023gor}%
  \BibitemOpen
  \bibfield  {author} {\bibinfo {author} {\bibfnamefont {G.}~\bibnamefont {Agazie}} \emph {et~al.} (\bibinfo {collaboration} {NANOGrav}),\ }\href {\doibase 10.3847/2041-8213/acdac6} {\bibfield  {journal} {\bibinfo  {journal} {Astrophys. J. Lett.}\ }\textbf {\bibinfo {volume} {951}},\ \bibinfo {pages} {L8} (\bibinfo {year} {2023}{\natexlab{b}})},\ \Eprint {http://arxiv.org/abs/2306.16213} {arXiv:2306.16213 [astro-ph.HE]} \BibitemShut {NoStop}%
\bibitem [{\citenamefont {Johnson}\ \emph {et~al.}(2024)\citenamefont {Johnson} \emph {et~al.}}]{NANOGrav:2023icp}%
  \BibitemOpen
  \bibfield  {author} {\bibinfo {author} {\bibfnamefont {A.~D.}\ \bibnamefont {Johnson}} \emph {et~al.} (\bibinfo {collaboration} {NANOGrav}),\ }\href {\doibase 10.1103/PhysRevD.109.103012} {\bibfield  {journal} {\bibinfo  {journal} {Phys. Rev. D}\ }\textbf {\bibinfo {volume} {109}},\ \bibinfo {pages} {103012} (\bibinfo {year} {2024})},\ \Eprint {http://arxiv.org/abs/2306.16223} {arXiv:2306.16223 [astro-ph.HE]} \BibitemShut {NoStop}%
\bibitem [{\citenamefont {Smarra}\ \emph {et~al.}(2023)\citenamefont {Smarra} \emph {et~al.}}]{EuropeanPulsarTimingArray:2023egv}%
  \BibitemOpen
  \bibfield  {author} {\bibinfo {author} {\bibfnamefont {C.}~\bibnamefont {Smarra}} \emph {et~al.} (\bibinfo {collaboration} {European Pulsar Timing Array}),\ }\href {\doibase 10.1103/PhysRevLett.131.171001} {\bibfield  {journal} {\bibinfo  {journal} {Phys. Rev. Lett.}\ }\textbf {\bibinfo {volume} {131}},\ \bibinfo {pages} {171001} (\bibinfo {year} {2023})},\ \Eprint {http://arxiv.org/abs/2306.16228} {arXiv:2306.16228 [astro-ph.HE]} \BibitemShut {NoStop}%
\bibitem [{\citenamefont {Antoniadis}\ \emph {et~al.}(2023)\citenamefont {Antoniadis} \emph {et~al.}}]{EPTA:2023fyk}%
  \BibitemOpen
  \bibfield  {author} {\bibinfo {author} {\bibfnamefont {J.}~\bibnamefont {Antoniadis}} \emph {et~al.} (\bibinfo {collaboration} {EPTA, InPTA:}),\ }\href {\doibase 10.1051/0004-6361/202346844} {\bibfield  {journal} {\bibinfo  {journal} {Astron. Astrophys.}\ }\textbf {\bibinfo {volume} {678}},\ \bibinfo {pages} {A50} (\bibinfo {year} {2023})},\ \Eprint {http://arxiv.org/abs/2306.16214} {arXiv:2306.16214 [astro-ph.HE]} \BibitemShut {NoStop}%
\bibitem [{\citenamefont {Quelquejay~Leclere}\ \emph {et~al.}(2023)\citenamefont {Quelquejay~Leclere} \emph {et~al.}}]{EuropeanPulsarTimingArray:2023lqe}%
  \BibitemOpen
  \bibfield  {author} {\bibinfo {author} {\bibfnamefont {H.}~\bibnamefont {Quelquejay~Leclere}} \emph {et~al.} (\bibinfo {collaboration} {European Pulsar Timing Array, EPTA}),\ }\href {\doibase 10.1103/PhysRevD.108.123527} {\bibfield  {journal} {\bibinfo  {journal} {Phys. Rev. D}\ }\textbf {\bibinfo {volume} {108}},\ \bibinfo {pages} {123527} (\bibinfo {year} {2023})},\ \Eprint {http://arxiv.org/abs/2306.12234} {arXiv:2306.12234 [gr-qc]} \BibitemShut {NoStop}%
\bibitem [{\citenamefont {Payne}\ \emph {et~al.}(2019)\citenamefont {Payne}, \citenamefont {Talbot},\ and\ \citenamefont {Thrane}}]{Payne:2019wmy}%
  \BibitemOpen
  \bibfield  {author} {\bibinfo {author} {\bibfnamefont {E.}~\bibnamefont {Payne}}, \bibinfo {author} {\bibfnamefont {C.}~\bibnamefont {Talbot}}, \ and\ \bibinfo {author} {\bibfnamefont {E.}~\bibnamefont {Thrane}},\ }\href {\doibase 10.1103/PhysRevD.100.123017} {\bibfield  {journal} {\bibinfo  {journal} {Phys. Rev. D}\ }\textbf {\bibinfo {volume} {100}},\ \bibinfo {pages} {123017} (\bibinfo {year} {2019})},\ \Eprint {http://arxiv.org/abs/1905.05477} {arXiv:1905.05477 [astro-ph.IM]} \BibitemShut {NoStop}%
\bibitem [{\citenamefont {Dickey}(1971)}]{Dickey1971TheWL}%
  \BibitemOpen
  \bibfield  {author} {\bibinfo {author} {\bibfnamefont {J.~M.}\ \bibnamefont {Dickey}},\ }\href {https://api.semanticscholar.org/CorpusID:123029751} {\bibfield  {journal} {\bibinfo  {journal} {Annals of Mathematical Statistics}\ }\textbf {\bibinfo {volume} {42}},\ \bibinfo {pages} {204} (\bibinfo {year} {1971})}\BibitemShut {NoStop}%
\bibitem [{\citenamefont {Garg}\ \emph {et~al.}(2024{\natexlab{b}})\citenamefont {Garg}, \citenamefont {Derdzinski}, \citenamefont {Tiwari}, \citenamefont {Gair},\ and\ \citenamefont {Mayer}}]{Garg:2024oeu}%
  \BibitemOpen
  \bibfield  {author} {\bibinfo {author} {\bibfnamefont {M.}~\bibnamefont {Garg}}, \bibinfo {author} {\bibfnamefont {A.}~\bibnamefont {Derdzinski}}, \bibinfo {author} {\bibfnamefont {S.}~\bibnamefont {Tiwari}}, \bibinfo {author} {\bibfnamefont {J.}~\bibnamefont {Gair}}, \ and\ \bibinfo {author} {\bibfnamefont {L.}~\bibnamefont {Mayer}},\ }\href {\doibase 10.1093/mnras/stae1764} {\bibfield  {journal} {\bibinfo  {journal} {Mon. Not. Roy. Astron. Soc.}\ }\textbf {\bibinfo {volume} {532}},\ \bibinfo {pages} {4060} (\bibinfo {year} {2024}{\natexlab{b}})},\ \Eprint {http://arxiv.org/abs/2402.14058} {arXiv:2402.14058 [astro-ph.GA]} \BibitemShut {NoStop}%
\bibitem [{\citenamefont {Chandramouli}\ \emph {et~al.}(2024)\citenamefont {Chandramouli}, \citenamefont {Prokup}, \citenamefont {Berti},\ and\ \citenamefont {Yunes}}]{Chandramouli:2024vhw}%
  \BibitemOpen
  \bibfield  {author} {\bibinfo {author} {\bibfnamefont {R.~S.}\ \bibnamefont {Chandramouli}}, \bibinfo {author} {\bibfnamefont {K.}~\bibnamefont {Prokup}}, \bibinfo {author} {\bibfnamefont {E.}~\bibnamefont {Berti}}, \ and\ \bibinfo {author} {\bibfnamefont {N.}~\bibnamefont {Yunes}},\ }\href@noop {} {\  (\bibinfo {year} {2024})},\ \Eprint {http://arxiv.org/abs/2410.06254} {arXiv:2410.06254 [gr-qc]} \BibitemShut {NoStop}%
\bibitem [{\citenamefont {Vallisneri}(2008)}]{Vallisneri:2007ev}%
  \BibitemOpen
  \bibfield  {author} {\bibinfo {author} {\bibfnamefont {M.}~\bibnamefont {Vallisneri}},\ }\href {\doibase 10.1103/PhysRevD.77.042001} {\bibfield  {journal} {\bibinfo  {journal} {Phys. Rev. D}\ }\textbf {\bibinfo {volume} {77}},\ \bibinfo {pages} {042001} (\bibinfo {year} {2008})},\ \Eprint {http://arxiv.org/abs/gr-qc/0703086} {arXiv:gr-qc/0703086} \BibitemShut {NoStop}%
\bibitem [{\citenamefont {Tikhonov}\ and\ \citenamefont {Arsenin}(1977)}]{tikhonov1977solutions}%
  \BibitemOpen
  \bibfield  {author} {\bibinfo {author} {\bibfnamefont {A.~N.}\ \bibnamefont {Tikhonov}}\ and\ \bibinfo {author} {\bibfnamefont {V.~Y.}\ \bibnamefont {Arsenin}},\ }\href@noop {} {\emph {\bibinfo {title} {Solutions of ill-posed problems}}}\ (\bibinfo  {publisher} {V. H. Winston \& Sons},\ \bibinfo {address} {Washington, D.C.: John Wiley \& Sons, New York},\ \bibinfo {year} {1977})\ pp.\ \bibinfo {pages} {xiii+258},\ \bibinfo {note} {translated from the Russian, Preface by translation editor Fritz John, Scripta Series in Mathematics}\BibitemShut {NoStop}%
\bibitem [{\citenamefont {Hoerl}\ and\ \citenamefont {Kennard}(1970)}]{481c956f-fc76-3a47-912c-132550d4ccb6}%
  \BibitemOpen
  \bibfield  {author} {\bibinfo {author} {\bibfnamefont {A.~E.}\ \bibnamefont {Hoerl}}\ and\ \bibinfo {author} {\bibfnamefont {R.~W.}\ \bibnamefont {Kennard}},\ }\href {http://www.jstor.org/stable/1267352} {\bibfield  {journal} {\bibinfo  {journal} {Technometrics}\ }\textbf {\bibinfo {volume} {12}},\ \bibinfo {pages} {69} (\bibinfo {year} {1970})}\BibitemShut {NoStop}%
\bibitem [{\citenamefont {Khalvati}\ \emph {et~al.}(2024)\citenamefont {Khalvati}, \citenamefont {Santini}, \citenamefont {Duque}, \citenamefont {Speri}, \citenamefont {Gair}, \citenamefont {Yang},\ and\ \citenamefont {Brito}}]{Khalvati:2024tzz}%
  \BibitemOpen
  \bibfield  {author} {\bibinfo {author} {\bibfnamefont {H.}~\bibnamefont {Khalvati}}, \bibinfo {author} {\bibfnamefont {A.}~\bibnamefont {Santini}}, \bibinfo {author} {\bibfnamefont {F.}~\bibnamefont {Duque}}, \bibinfo {author} {\bibfnamefont {L.}~\bibnamefont {Speri}}, \bibinfo {author} {\bibfnamefont {J.}~\bibnamefont {Gair}}, \bibinfo {author} {\bibfnamefont {H.}~\bibnamefont {Yang}}, \ and\ \bibinfo {author} {\bibfnamefont {R.}~\bibnamefont {Brito}},\ }\href@noop {} {\  (\bibinfo {year} {2024})},\ \Eprint {http://arxiv.org/abs/2410.17310} {arXiv:2410.17310 [gr-qc]} \BibitemShut {NoStop}%
\bibitem [{\citenamefont {Katz}\ \emph {et~al.}(2021)\citenamefont {Katz}, \citenamefont {Chua}, \citenamefont {Speri}, \citenamefont {Warburton},\ and\ \citenamefont {Hughes}}]{Katz:2021yft}%
  \BibitemOpen
  \bibfield  {author} {\bibinfo {author} {\bibfnamefont {M.~L.}\ \bibnamefont {Katz}}, \bibinfo {author} {\bibfnamefont {A.~J.~K.}\ \bibnamefont {Chua}}, \bibinfo {author} {\bibfnamefont {L.}~\bibnamefont {Speri}}, \bibinfo {author} {\bibfnamefont {N.}~\bibnamefont {Warburton}}, \ and\ \bibinfo {author} {\bibfnamefont {S.~A.}\ \bibnamefont {Hughes}},\ }\href {\doibase 10.1103/PhysRevD.104.064047} {\bibfield  {journal} {\bibinfo  {journal} {Phys. Rev. D}\ }\textbf {\bibinfo {volume} {104}},\ \bibinfo {pages} {064047} (\bibinfo {year} {2021})},\ \Eprint {http://arxiv.org/abs/2104.04582} {arXiv:2104.04582 [gr-qc]} \BibitemShut {NoStop}%
\bibitem [{\citenamefont {Chua}\ \emph {et~al.}(2021)\citenamefont {Chua}, \citenamefont {Katz}, \citenamefont {Warburton},\ and\ \citenamefont {Hughes}}]{Chua:2020stf}%
  \BibitemOpen
  \bibfield  {author} {\bibinfo {author} {\bibfnamefont {A.~J.~K.}\ \bibnamefont {Chua}}, \bibinfo {author} {\bibfnamefont {M.~L.}\ \bibnamefont {Katz}}, \bibinfo {author} {\bibfnamefont {N.}~\bibnamefont {Warburton}}, \ and\ \bibinfo {author} {\bibfnamefont {S.~A.}\ \bibnamefont {Hughes}},\ }\href {\doibase 10.1103/PhysRevLett.126.051102} {\bibfield  {journal} {\bibinfo  {journal} {Phys. Rev. Lett.}\ }\textbf {\bibinfo {volume} {126}},\ \bibinfo {pages} {051102} (\bibinfo {year} {2021})},\ \Eprint {http://arxiv.org/abs/2008.06071} {arXiv:2008.06071 [gr-qc]} \BibitemShut {NoStop}%
\bibitem [{\citenamefont {Karnesis}\ \emph {et~al.}(2023)\citenamefont {Karnesis}, \citenamefont {Katz}, \citenamefont {Korsakova}, \citenamefont {Gair},\ and\ \citenamefont {Stergioulas}}]{Karnesis:2023ras}%
  \BibitemOpen
  \bibfield  {author} {\bibinfo {author} {\bibfnamefont {N.}~\bibnamefont {Karnesis}}, \bibinfo {author} {\bibfnamefont {M.~L.}\ \bibnamefont {Katz}}, \bibinfo {author} {\bibfnamefont {N.}~\bibnamefont {Korsakova}}, \bibinfo {author} {\bibfnamefont {J.~R.}\ \bibnamefont {Gair}}, \ and\ \bibinfo {author} {\bibfnamefont {N.}~\bibnamefont {Stergioulas}},\ }\href {\doibase 10.1093/mnras/stad2939} {\bibfield  {journal} {\bibinfo  {journal} {Mon. Not. Roy. Astron. Soc.}\ }\textbf {\bibinfo {volume} {526}},\ \bibinfo {pages} {4814} (\bibinfo {year} {2023})},\ \Eprint {http://arxiv.org/abs/2303.02164} {arXiv:2303.02164 [astro-ph.IM]} \BibitemShut {NoStop}%
\bibitem [{\citenamefont {Katz}\ \emph {et~al.}(2024{\natexlab{a}})\citenamefont {Katz}, \citenamefont {Chapman-Bird}, \citenamefont {Speri}, \citenamefont {Karnesis},\ and\ \citenamefont {Korsakova}}]{michael_katz_2024_10930980}%
  \BibitemOpen
  \bibfield  {author} {\bibinfo {author} {\bibfnamefont {M.}~\bibnamefont {Katz}}, \bibinfo {author} {\bibfnamefont {C.}~\bibnamefont {Chapman-Bird}}, \bibinfo {author} {\bibfnamefont {L.}~\bibnamefont {Speri}}, \bibinfo {author} {\bibfnamefont {N.}~\bibnamefont {Karnesis}}, \ and\ \bibinfo {author} {\bibfnamefont {N.}~\bibnamefont {Korsakova}},\ }\href {\doibase 10.5281/zenodo.10930980} {\enquote {\bibinfo {title} {mikekatz04/lisaanalysistools: First main release.}}\ } (\bibinfo {year} {2024}{\natexlab{a}})\BibitemShut {NoStop}%
\bibitem [{\citenamefont {Katz}\ \emph {et~al.}(2022)\citenamefont {Katz}, \citenamefont {Bayle}, \citenamefont {Chua},\ and\ \citenamefont {Vallisneri}}]{Katz:2022yqe}%
  \BibitemOpen
  \bibfield  {author} {\bibinfo {author} {\bibfnamefont {M.~L.}\ \bibnamefont {Katz}}, \bibinfo {author} {\bibfnamefont {J.-B.}\ \bibnamefont {Bayle}}, \bibinfo {author} {\bibfnamefont {A.~J.~K.}\ \bibnamefont {Chua}}, \ and\ \bibinfo {author} {\bibfnamefont {M.}~\bibnamefont {Vallisneri}},\ }\href {\doibase 10.1103/PhysRevD.106.103001} {\bibfield  {journal} {\bibinfo  {journal} {Phys. Rev. D}\ }\textbf {\bibinfo {volume} {106}},\ \bibinfo {pages} {103001} (\bibinfo {year} {2022})},\ \Eprint {http://arxiv.org/abs/2204.06633} {arXiv:2204.06633 [gr-qc]} \BibitemShut {NoStop}%
\bibitem [{\citenamefont {Kejriwal}\ \emph {et~al.}(tion)\citenamefont {Kejriwal}, \citenamefont {Burke},\ and\ \citenamefont {Chapman-Bird}}]{kejriwal_2024_sef}%
  \BibitemOpen
  \bibfield  {author} {\bibinfo {author} {\bibfnamefont {S.}~\bibnamefont {Kejriwal}}, \bibinfo {author} {\bibfnamefont {O.}~\bibnamefont {Burke}}, \ and\ \bibinfo {author} {\bibfnamefont {C.}~\bibnamefont {Chapman-Bird}},\ }\href {https://github.com/perturber/StableEMRIFisher} {\enquote {\bibinfo {title} {Stableemrifisher (sef)},}\ } (\bibinfo {year} {manuscript in preparation})\BibitemShut {NoStop}%
\bibitem [{\citenamefont {Saleh}\ \emph {et~al.}(2024)\citenamefont {Saleh}, \citenamefont {Zimmerman}, \citenamefont {Chen},\ and\ \citenamefont {Ghattas}}]{Saleh:2024tgr}%
  \BibitemOpen
  \bibfield  {author} {\bibinfo {author} {\bibfnamefont {B.}~\bibnamefont {Saleh}}, \bibinfo {author} {\bibfnamefont {A.}~\bibnamefont {Zimmerman}}, \bibinfo {author} {\bibfnamefont {P.}~\bibnamefont {Chen}}, \ and\ \bibinfo {author} {\bibfnamefont {O.}~\bibnamefont {Ghattas}},\ }\href {\doibase 10.1103/PhysRevD.110.104037} {\bibfield  {journal} {\bibinfo  {journal} {Phys. Rev. D}\ }\textbf {\bibinfo {volume} {110}},\ \bibinfo {pages} {104037} (\bibinfo {year} {2024})},\ \Eprint {http://arxiv.org/abs/2405.19407} {arXiv:2405.19407 [gr-qc]} \BibitemShut {NoStop}%
\bibitem [{\citenamefont {Gupta}\ \emph {et~al.}(2024)\citenamefont {Gupta} \emph {et~al.}}]{Gupta:2024gun}%
  \BibitemOpen
  \bibfield  {author} {\bibinfo {author} {\bibfnamefont {A.}~\bibnamefont {Gupta}} \emph {et~al.},\ }\href@noop {} {\  (\bibinfo {year} {2024})},\ \Eprint {http://arxiv.org/abs/2405.02197} {arXiv:2405.02197 [gr-qc]} \BibitemShut {NoStop}%
\bibitem [{\citenamefont {Li}\ \emph {et~al.}(2012)\citenamefont {Li}, \citenamefont {Del~Pozzo}, \citenamefont {Vitale}, \citenamefont {Van Den~Broeck}, \citenamefont {Agathos}, \citenamefont {Veitch}, \citenamefont {Grover}, \citenamefont {Sidery}, \citenamefont {Sturani},\ and\ \citenamefont {Vecchio}}]{Li:2011cg}%
  \BibitemOpen
  \bibfield  {author} {\bibinfo {author} {\bibfnamefont {T.~G.~F.}\ \bibnamefont {Li}}, \bibinfo {author} {\bibfnamefont {W.}~\bibnamefont {Del~Pozzo}}, \bibinfo {author} {\bibfnamefont {S.}~\bibnamefont {Vitale}}, \bibinfo {author} {\bibfnamefont {C.}~\bibnamefont {Van Den~Broeck}}, \bibinfo {author} {\bibfnamefont {M.}~\bibnamefont {Agathos}}, \bibinfo {author} {\bibfnamefont {J.}~\bibnamefont {Veitch}}, \bibinfo {author} {\bibfnamefont {K.}~\bibnamefont {Grover}}, \bibinfo {author} {\bibfnamefont {T.}~\bibnamefont {Sidery}}, \bibinfo {author} {\bibfnamefont {R.}~\bibnamefont {Sturani}}, \ and\ \bibinfo {author} {\bibfnamefont {A.}~\bibnamefont {Vecchio}},\ }\href {\doibase 10.1103/PhysRevD.85.082003} {\bibfield  {journal} {\bibinfo  {journal} {Phys. Rev. D}\ }\textbf {\bibinfo {volume} {85}},\ \bibinfo {pages} {082003} (\bibinfo {year} {2012})},\ \Eprint {http://arxiv.org/abs/1110.0530} {arXiv:1110.0530 [gr-qc]} \BibitemShut {NoStop}%
\bibitem [{\citenamefont {Saleem}\ \emph {et~al.}(2022)\citenamefont {Saleem}, \citenamefont {Datta}, \citenamefont {Arun},\ and\ \citenamefont {Sathyaprakash}}]{Saleem:2021nsb}%
  \BibitemOpen
  \bibfield  {author} {\bibinfo {author} {\bibfnamefont {M.}~\bibnamefont {Saleem}}, \bibinfo {author} {\bibfnamefont {S.}~\bibnamefont {Datta}}, \bibinfo {author} {\bibfnamefont {K.~G.}\ \bibnamefont {Arun}}, \ and\ \bibinfo {author} {\bibfnamefont {B.~S.}\ \bibnamefont {Sathyaprakash}},\ }\href {\doibase 10.1103/PhysRevD.105.084062} {\bibfield  {journal} {\bibinfo  {journal} {Phys. Rev. D}\ }\textbf {\bibinfo {volume} {105}},\ \bibinfo {pages} {084062} (\bibinfo {year} {2022})},\ \Eprint {http://arxiv.org/abs/2110.10147} {arXiv:2110.10147 [gr-qc]} \BibitemShut {NoStop}%
\bibitem [{\citenamefont {Chua}\ and\ \citenamefont {Vallisneri}(2020)}]{Chua:2020oxn}%
  \BibitemOpen
  \bibfield  {author} {\bibinfo {author} {\bibfnamefont {A.~J.~K.}\ \bibnamefont {Chua}}\ and\ \bibinfo {author} {\bibfnamefont {M.}~\bibnamefont {Vallisneri}},\ }\href@noop {} {\  (\bibinfo {year} {2020})},\ \Eprint {http://arxiv.org/abs/2006.08918} {arXiv:2006.08918 [gr-qc]} \BibitemShut {NoStop}%
\bibitem [{\citenamefont {Vallisneri}(2009)}]{Vallisneri:2008ye}%
  \BibitemOpen
  \bibfield  {author} {\bibinfo {author} {\bibfnamefont {M.}~\bibnamefont {Vallisneri}},\ }\href {\doibase 10.1088/0264-9381/26/9/094024} {\bibfield  {journal} {\bibinfo  {journal} {Class. Quant. Grav.}\ }\textbf {\bibinfo {volume} {26}},\ \bibinfo {pages} {094024} (\bibinfo {year} {2009})},\ \Eprint {http://arxiv.org/abs/0812.0751} {arXiv:0812.0751 [gr-qc]} \BibitemShut {NoStop}%
\bibitem [{\citenamefont {Littenberg}\ and\ \citenamefont {Cornish}(2023)}]{Littenberg:2023xpl}%
  \BibitemOpen
  \bibfield  {author} {\bibinfo {author} {\bibfnamefont {T.~B.}\ \bibnamefont {Littenberg}}\ and\ \bibinfo {author} {\bibfnamefont {N.~J.}\ \bibnamefont {Cornish}},\ }\href {\doibase 10.1103/PhysRevD.107.063004} {\bibfield  {journal} {\bibinfo  {journal} {Phys. Rev. D}\ }\textbf {\bibinfo {volume} {107}},\ \bibinfo {pages} {063004} (\bibinfo {year} {2023})},\ \Eprint {http://arxiv.org/abs/2301.03673} {arXiv:2301.03673 [gr-qc]} \BibitemShut {NoStop}%
\bibitem [{\citenamefont {Katz}\ \emph {et~al.}(2024{\natexlab{b}})\citenamefont {Katz}, \citenamefont {Karnesis}, \citenamefont {Korsakova}, \citenamefont {Gair},\ and\ \citenamefont {Stergioulas}}]{Katz:2024oqg}%
  \BibitemOpen
  \bibfield  {author} {\bibinfo {author} {\bibfnamefont {M.~L.}\ \bibnamefont {Katz}}, \bibinfo {author} {\bibfnamefont {N.}~\bibnamefont {Karnesis}}, \bibinfo {author} {\bibfnamefont {N.}~\bibnamefont {Korsakova}}, \bibinfo {author} {\bibfnamefont {J.~R.}\ \bibnamefont {Gair}}, \ and\ \bibinfo {author} {\bibfnamefont {N.}~\bibnamefont {Stergioulas}},\ }\href@noop {} {\  (\bibinfo {year} {2024}{\natexlab{b}})},\ \Eprint {http://arxiv.org/abs/2405.04690} {arXiv:2405.04690 [gr-qc]} \BibitemShut {NoStop}%
\bibitem [{\citenamefont {Green}(1995)}]{Green:1995mxx}%
  \BibitemOpen
  \bibfield  {author} {\bibinfo {author} {\bibfnamefont {P.~J.}\ \bibnamefont {Green}},\ }\href {\doibase 10.1093/biomet/82.4.711} {\bibfield  {journal} {\bibinfo  {journal} {Biometrika}\ }\textbf {\bibinfo {volume} {82}},\ \bibinfo {pages} {711} (\bibinfo {year} {1995})}\BibitemShut {NoStop}%
\bibitem [{\citenamefont {Gelfand}\ and\ \citenamefont {Smith}(1990)}]{f47e9344-75dc-3782-a811-9e51a952e8bf}%
  \BibitemOpen
  \bibfield  {author} {\bibinfo {author} {\bibfnamefont {A.~E.}\ \bibnamefont {Gelfand}}\ and\ \bibinfo {author} {\bibfnamefont {A.~F.~M.}\ \bibnamefont {Smith}},\ }\href {http://www.jstor.org/stable/2289776} {\bibfield  {journal} {\bibinfo  {journal} {Journal of the American Statistical Association}\ }\textbf {\bibinfo {volume} {85}},\ \bibinfo {pages} {398} (\bibinfo {year} {1990})}\BibitemShut {NoStop}%
\bibitem [{\citenamefont {Liu}(2008)}]{10.5555/1571802}%
  \BibitemOpen
  \bibfield  {author} {\bibinfo {author} {\bibfnamefont {J.~S.}\ \bibnamefont {Liu}},\ }\href@noop {} {\emph {\bibinfo {title} {Monte Carlo Strategies in Scientific Computing}}}\ (\bibinfo  {publisher} {Springer Publishing Company, Incorporated},\ \bibinfo {year} {2008})\BibitemShut {NoStop}%
\end{thebibliography}%

\end{document}